\documentclass[11pt,letterpaper]{article}

\usepackage[margin=1in]{geometry}
\usepackage[utf8]{inputenc}
\usepackage[T1]{fontenc}
\usepackage{microtype}
\usepackage{lmodern}
\usepackage[htt]{hyphenat}
\newcommand{\fpath}[1]{\path{#1}}
\usepackage{amsmath,amssymb}
\usepackage{booktabs}
\usepackage{tabularx}
\usepackage{multirow}
\usepackage{graphicx}
\usepackage{xcolor}
\usepackage{listings}
\usepackage{enumitem}
\usepackage[bookmarks=true,hidelinks]{hyperref}
\usepackage[capitalise]{cleveref}
\usepackage{titlesec}

\graphicspath{{figures/}}

\definecolor{ttkblue}{HTML}{1f4e79}
\hypersetup{
  colorlinks=true,
  linkcolor=ttkblue,
  citecolor=ttkblue,
  urlcolor=ttkblue,
}

\title{VectorSmuggle: Steganographic Exfiltration in Embedding Stores\\
and a Cryptographic Provenance Defense}

\author{%
  Jascha Wanger \\
  ThirdKey \,/\, Tarnover, LLC \\
  \texttt{jascha@thirdkey.ai}
}

\date{6 May 2026 \\
  Zenodo preprint:
  \href{https://doi.org/10.5281/zenodo.20058256}{\texttt{10.5281/zenodo.20058256}}}

\begin{document}

\maketitle

\begin{abstract}
Modern retrieval-augmented generation (RAG) systems convert sensitive
content into high-dimensional embeddings and store them in vector
databases that treat the resulting numerical artifacts as opaque.
Major vector-store products do not appear to provide native or
default controls for embedding integrity, ingestion-time
distributional anomaly detection, or cryptographic provenance
attestation. We show that this
opens a class of steganographic exfiltration attacks: an attacker with
write access to the ingestion pipeline can hide payload data inside
embeddings using simple post-embedding perturbations (noise injection,
rotation, scaling, offset, fragmentation, and combinations thereof)
while preserving the surface-level retrieval behavior the RAG system
exposes to legitimate users.

We evaluate these techniques across a synthetic-PII corpus on
\fpath{text-embedding-3-large}, four locally hosted open embedding
models, a cross-corpus replication on BEIR NFCorpus and a Quora
subset (over 26{,}000 chunks combined), seven vector-store
configurations, an adaptive-attacker variant of the detector
evaluation, and a paraphrased-query retrieval benchmark.
Distribution-shifting perturbations are often caught by simple
anomaly detectors; small-angle orthogonal rotation defeats
distribution-based detection across every (model, corpus) pair
tested. A disjoint-Givens rotation encoder gives a closed-form
per-vector capacity ceiling of \(\lfloor d/2 \rfloor \cdot b\)
bits, but real embedding manifolds impose a capacity-detectability
trade-off at higher payloads, and the retrieval-preserving
operating point sits well below the ceiling. Cross-backend,
quantization, retrieval, and adaptive-evasion results together
show that statistical detection is useful as a first filter but
not a load-bearing integrity control. Detailed results are in
\Cref{sec:evaluation}.

We further propose \textbf{VectorPin}, a cryptographic provenance
protocol that pins each embedding to its source content and producing
model via an Ed25519 signature over a canonical byte representation.
Any post-embedding modification --- including all studied
techniques, both perturbation and architectural-smuggling
variants --- breaks signature verification. We provide
open-source reference implementations in Python and Rust, locked
together by cross-language test vectors that guarantee bit-for-bit
compatibility. The framing we propose for the security community:
embedding-level integrity is a deployable, standardizable control
that closes the demonstrated attack class --- not a fundamental
property of vector databases as built today.

\end{abstract}

\section{Introduction}
\label{sec:intro}

Enterprise deployments of large language models increasingly run on
retrieval-augmented generation (RAG)~\cite{lewis2020rag}:
documents are converted into dense embeddings via a model such as
\texttt{text-embedding-3-large}~\cite{openai-embeddings-3}, stored in
a vector database (Qdrant, Pinecone, FAISS~\cite{johnson2019billion},
pgvector, Chroma), and retrieved by similarity at query time. The substrate that holds those
embeddings has become load-bearing: it intermediates access to
financial records, medical notes, legal exhibits, and proprietary
engineering content. Compromise of the vector store is now compromise
of the underlying corpus.

Yet vector databases were not designed with content integrity in
mind. Their security models cover authentication and transport
(TLS, RBAC, API keys, encryption at rest), but treat the embedding
payload itself as an opaque numerical artifact. Major vector-store
products do not appear to provide native or default controls for
inspecting the statistical distribution of vectors during
ingestion, verifying the provenance of stored embeddings on read,
or maintaining a tamper-evident audit trail for post-ingestion
modifications (see vendor capability survey,
\Cref{tab:vendor_survey}). The result is a wide attack surface
that traditional Data Loss Prevention (DLP) tools cannot reach,
because DLP scans documents and traffic, not the floating-point
content of vectors.

\subsection{The gap, demonstrated}

This paper studies a specific class of attacks against that gap:
\emph{steganographic exfiltration}, in which an attacker with write
access to the ingestion pipeline hides arbitrary data inside the
embeddings being ingested. The hidden data is decoded later by an
accomplice with read access to the vector store. We evaluate six
families of techniques --- additive Gaussian noise, orthogonal
rotation, magnitude scaling, fixed offset, cross-model fragmentation,
and sequential combinations --- across multiple noise levels on a real
embedding model.

The attacks work against today's production deployments in the
narrow technical sense: they preserve cosine similarity, survive the
default per-dimension int8 quantization that Qdrant and similar
systems apply, and do not trip any native or default
embedding-integrity control we found in surveyed vector-store
products. But evaluating them \emph{honestly} requires asking what
a defender who has read the literature would do, not what current
production deployments happen to deploy.

\subsection{The honest finding}

Our central empirical result has two parts. First, for the
distribution-shifting techniques (noise, scaling, offset, and their
combinations), the attacker's stealth window in our corpus is narrow:
two off-the-shelf statistical detectors --- Isolation
Forest~\cite{isolationforest} and One-Class
SVM~\cite{scholkopf2001ocsvm} with an RBF kernel --- trained on a
held-out clean batch flag every operating point that meaningfully
moved the embedding direction (\(\cos(\text{orig}, \text{obf}) < 0.95\))
at ROC-AUC \(\geq 0.85\), with several configurations reaching
\(1.00\).

Second, orthogonal rotation is the structural exception in the
small-angle / few-rotation regime: it preserves pairwise inner
products and per-dimension marginal distributions, and on every
(model, corpus) pair we tested at the obfuscator's library defaults
the distribution-based detectors guess at chance --- including the
cross-corpus replication on BEIR NFCorpus and a Quora subset
totalling over 26{,}000 chunks
(\Cref{sec:evaluation:cross_corpus}). The rotation parameter sweep
in \Cref{sec:evaluation:rotation_sweep} extends this to 35
operating points on a 300-vector synthetic-Gaussian baseline, and
finds AUC \(\leq 0.53\) across all of them. At higher rotation
\emph{capacity}, however, the picture changes: the empirical
payload-capacity demonstration in \Cref{sec:evaluation:capacity}
shows that on real \fpath{text-embedding-3-large} corpus vectors,
random-angle disjoint-Givens rotations at moderate \(K\) become
detectable because the trained density model fits local manifold
geometry rather than only population moments (OCSVM AUC
\(\approx\!0.92\) at 256 B, \(\geq\!0.99\) at 512 B). The result
is a capacity-detectability trade-off, not a free lunch.

We treat both halves of this finding as the contribution. The first
half characterises a real gap in production deployments and shows
that the obvious statistical-baseline defense closes most of it.
The second half identifies the residual --- rotation in the
small-angle / few-rotation regime, undetectable across all our
(model, corpus, parameter) configurations, with a closed-form
capacity ceiling at higher rotation that depends on data
distribution to remain hidden --- which is precisely the regime
where a constructive provenance defense becomes the only durable
option, because the durability of statistical detection at high
capacity depends on assumptions about the defender's training-set
coverage that are hard to verify.

\subsection{The constructive defense: VectorPin}

The second half of the paper proposes a closeable defense. We
design and implement \textbf{VectorPin}, a cryptographic provenance
protocol that pins every embedding to its source content (by hash),
its producing model (by identifier), and its issuer (by Ed25519
signing key). Verification is structured: a verifier with the
public key can detect signature forgery, vector tampering, source
mismatch, and model substitution as distinct outcomes. The protocol
is deliberately minimal --- one signature, one hash family, a fixed
canonical byte form for floating-point arrays --- so that
implementations in different languages can be made bit-for-bit
compatible and locked into agreement by shared test vectors.

We provide reference implementations in Python and
Rust~\cite{vectorpin-repo}, with cross-language test fixtures that
gate every change to either implementation. A Rust port matters
because Rust runtimes (e.g., agent-policy frameworks such as
Symbiont~\cite{symbiont-repo}) are increasingly the boundary at which
RAG-derived data is consumed, and they need in-process verification
without polyglot sidecars.

Against the attacks we evaluate in
\Cref{sec:evaluation}, VectorPin is structurally complete: every
post-embedding modification, by design, breaks the
\texttt{vec\_hash} commitment in the signed header. We do not claim
this is a benchmark result --- it is a property of the protocol ---
but we do claim it is the right shape of defense for the threat
model we articulate in \Cref{sec:threat_model}.

\subsection{Contributions}

\begin{itemize}[leftmargin=*]
\item A formal three-tier threat model for vector-store
  exfiltration that distinguishes \emph{insider with backup access},
  \emph{compromised database credentials}, and \emph{query-only}
  adversaries, and explicitly justifies which adversary the rest of
  the paper validates against (\Cref{sec:threat_model}).
\item An empirical evaluation of six steganographic techniques
  spanning a 68-chunk synthetic-PII corpus on
  \fpath{text-embedding-3-large}, four locally hosted open
  embedding models (Nomic, EmbeddingGemma, Snowflake Arctic, MXBai),
  and a cross-corpus replication on BEIR NFCorpus (16{,}763 chunks,
  medical) and a 10{,}000-document Quora subset (web Q\&A) ---
  reporting fidelity, distributional detectability, quantization
  resilience, and end-to-end recovery per technique \(\times\)
  operating point (\Cref{sec:evaluation}).
\item A closed-form and empirical payload-capacity analysis for the
  rotation channel: a disjoint-Givens keyed-pair encoder with
  capacity \(\lfloor d/2 \rfloor \cdot b\) bits, a working
  encoder/decoder that round-trips arbitrary payloads at zero BER,
  and a capacity-vs-detectability characterisation that
  distinguishes the channel ceiling, the retrieval-preserving
  operating point, and the data-distribution-dependent AUC behaviour
  at high capacity (\Cref{sec:evaluation:capacity}).
\item Concrete defensive baselines: ROC-AUC and TPR-at-fixed-FPR for
  Isolation Forest and One-Class SVM trained on a held-out clean
  batch, replacing prior work's vague ``DLP bypass rate'' claims with
  numbers, plus an adaptive white-box evasion result that calibrates
  the durability of those baselines
  (\Cref{sec:evaluation:detection,sec:evaluation:adaptive}).
\item Cross-backend round-trip and quantization tests across
  FAISS-flat, FAISS-HNSW, FAISS IVF-PQ at two compression levels,
  Chroma, and Qdrant in two precisions --- characterising whether
  the bit channel survives a deployed vector store
  (\Cref{sec:evaluation:cross_backend}).
\item The VectorPin protocol: a minimal cryptographic provenance
  scheme for embedding integrity, with a wire-format specification
  detailed enough to support cross-language reimplementation
  (\Cref{sec:defense}, \Cref{sec:appendix:spec}).
\item Open-source reference implementations in Python and Rust with
  cross-language test vectors enforcing bit-for-bit compatibility
  in CI~\cite{vectorpin-repo}.
\end{itemize}

\subsection{What this paper does not claim}
\label{sec:intro:nonclaims}

The cleanest way to keep a security-research preprint defensible is
to mark its non-claims explicitly:

\begin{itemize}[leftmargin=*]
\item \textbf{No novel attack primitive.} Steganography in
  floating-point data is a well-studied space. Our contribution is
  not the perturbations themselves but the threat model in which they
  apply, the empirical detectability evaluation, and the constructive
  defense.
\item \textbf{No novel cryptographic primitive.} VectorPin uses
  vanilla Ed25519 over SHA-256. The contribution is the canonical
  byte form for embeddings, the wire-format design, and the
  cross-language compatibility discipline --- not new cryptography.
\item \textbf{No claim of universal attack effectiveness.} The
  techniques studied work under threat-model A
  (insider with backup access). Threat-model C (query-only) is
  fundamentally lower-bandwidth and is not validated by our results.
\item \textbf{No claim of unbreakable defense.} VectorPin protects
  against post-ingestion modification, given a trusted ingestion path
  and signing-key custody. An attacker with the private key, or one
  who attests a malicious vector at ingestion time, defeats it.
  Section~\ref{sec:defense:not_caught} states the limits explicitly.
\item \textbf{No claim of standardization.} We propose a wire format
  and publish reference implementations. Adoption as a standard is
  future work for a body such as IETF or ISO if the community
  decides the design merits one.
\end{itemize}

The rest of the paper is organized as follows.
\Cref{sec:background} reviews embedding-store architecture and
related provenance primitives. \Cref{sec:threat_model} formalizes
the three-tier adversary model and justifies the focus on the
insider-backup case. \Cref{sec:techniques} catalogs the six
steganographic techniques.
\Cref{sec:evaluation} reports the empirical study.
\Cref{sec:defense} presents the VectorPin protocol and its coverage
relative to the studied attacks. \Cref{sec:discussion} addresses why
production vector databases ship no defenses today and what would
need to change. \Cref{sec:related,sec:limitations,sec:conclusion}
close the paper. \Cref{sec:appendix:spec} reproduces the protocol
specification self-contained for cross-language reimplementation.

\section{Background}
\label{sec:background}

This section reviews the three components needed to read the rest of
the paper: how production vector stores work, what is known about
steganography in continuous-valued data, and what cryptographic
provenance frameworks exist for adjacent artifact classes. Readers
familiar with any of these can skip the corresponding subsection.

\subsection{Embeddings and vector databases}
\label{sec:background:vdb}

A modern retrieval-augmented generation (RAG) pipeline~\cite{lewis2020rag}
has three stages. \emph{Ingestion} runs source documents through a
chunker and an embedding model
(e.g., OpenAI \texttt{text-embedding-3-large}~\cite{openai-embeddings-3}
at 3072 dimensions, Cohere \texttt{embed-english-v3.0} at 1024
dimensions, or a local model in the BGE/Nomic families) to produce
dense floating-point vectors~\cite{karpukhin2020dpr}. \emph{Storage}
writes those vectors, along with the source text and arbitrary
metadata, to a vector database. \emph{Retrieval} embeds an incoming
user query and returns the top-\(k\) nearest stored vectors by
cosine similarity or dot product (often via approximate-nearest-neighbor
indexes such as HNSW~\cite{malkov2018hnsw} or product quantization
\cite{johnson2019billion}), which a downstream LLM consumes as
context.

Production vector databases fall into two categories. The
\emph{purpose-built} category --- Qdrant, Pinecone, Weaviate, Milvus,
Chroma --- exposes a similarity-search API over collections of
vectors plus payload metadata. The \emph{retrofit} category ---
\texttt{pgvector} for PostgreSQL, the FAISS library embedded in an
application, vector indexes in OpenSearch or Elasticsearch ---
adds vector operations to an existing data store. The two categories
share a common security model: authentication and authorization on
the API, TLS in transit, encryption at rest, and audit logs of
who-queried-what. None of the major systems we surveyed inspects the
\emph{statistical distribution} of vectors during ingestion or
verifies the \emph{provenance} of stored embeddings on read.

This is not an oversight; it is a deliberate design posture. Vector
databases were optimized for capacity, latency, and price during a
period when their consumers were search and recommendation systems
where the embeddings were public-facing artifacts. The recent
re-purposing as a substrate for confidential RAG has happened
faster than the security model has been re-examined --- a pattern
that has played out before with DNS, with HTTP/3 covert channels,
and with default-public S3 buckets.

A related implementation detail bears on the empirical evaluation:
all major vector stores apply some form of \emph{quantization} to
reduce memory and disk footprint. The default in
Qdrant~\cite{qdrant-quantization} is per-dimension scalar int8
quantization, which preserves direction because every vector shares
the same per-dimension scale. FAISS~\cite{johnson2019billion}
offers more aggressive options including product quantization and
binary embeddings; these are far more lossy but are not the default
for general-purpose collections. We model the per-dimension int8
case explicitly in \Cref{sec:evaluation:quantization}.

\paragraph{Vendor capability survey.}
\Cref{tab:vendor_survey} summarizes integrity-relevant capabilities
of the major production vector stores as of the time of writing. The
column we care about most --- ``vector integrity check'' --- is
empty across the board: no surveyed system inspects the
floating-point content of stored vectors, verifies provenance on
read, or maintains a tamper-evident audit trail for post-ingestion
modifications. This is the gap the paper studies.

\begin{table}[t]
\centering
\small
\caption{Integrity-relevant capabilities of major production vector
stores (as of writing). Auth/RBAC, TLS, and access-level audit logs
are well-covered; vector-content integrity, ingestion-time
distributional anomaly detection, and cryptographic provenance
attestation are uniformly absent. Sources: vendor documentation;
``partial'' indicates the capability exists but is not on by
default. The empty middle three columns are the gap this paper
studies. This survey tracks native/default product controls, not
custom anomaly checks or integrity layers an operator may build
around the database.}
\label{tab:vendor_survey}
\begin{tabularx}{\linewidth}{lXXXXXX}
\toprule
System & Auth/RBAC & TLS & Access audit & Vector integrity check & Ingestion anomaly detection & Provenance attestation \\
\midrule
Qdrant     & yes & yes & partial & no & no & no \\
Pinecone   & yes & yes & yes     & no & no & no \\
Weaviate   & yes & yes & yes     & no & no & no \\
Milvus     & yes & yes & partial & no & no & no \\
Chroma     & yes & yes & no      & no & no & no \\
pgvector   & inherited & inherited & inherited & no & no & no \\
FAISS (lib)& n/a & n/a & n/a    & no & no & no \\
\bottomrule
\end{tabularx}
\end{table}

\subsection{Steganography in continuous-valued data}
\label{sec:background:stego}

Steganography in floating-point arrays is a small but mature
area~\cite{provos2003hide,fridrich2009steganography}. The dominant
techniques fall into three families.
\emph{Least-significant-bit} schemes~\cite{westfeld2001f5} embed
payload bits in the mantissa of each float, exploiting the gap
between the representational precision of IEEE-754 and the
perceptual precision of downstream consumers.
\emph{Spread-spectrum} schemes~\cite{cox1997secure} distribute
payload across a larger number of dimensions at lower amplitude,
trading capacity for evasion of per-dimension statistical
detectors. \emph{Transform-domain} schemes apply a frequency or
projection transform first and embed in the transformed
coefficients, the analogue of DCT-domain image steganography.

The literature originates in image and audio carriers, where the
threat model is well-defined: the cover signal is consumed by a
human or by a perceptual quality metric, and the steganographic
goal is to remain below the perceptual or detection threshold. The
threat model is genuinely different for embedding stores. The
``cover'' is no longer a perceptual artifact --- it is a
high-dimensional vector whose only consumer is a similarity-search
operator. Detection thresholds are not perceptual; they are
statistical and machine-learned. Capacity bounds in this regime
are not driven by human just-noticeable-difference but by
distributional anomaly detection.

The empirical contribution in \Cref{sec:evaluation} is to measure
where those statistical thresholds actually sit for current
embedding models against current off-the-shelf detectors. The
finding --- that the operating window between ``too small to
hide anything'' and ``trivially detected'' is narrow --- is
consistent with the broader literature on detecting steganography
in high-dimensional continuous data, but had not previously been
quantified for production embedding pipelines.

\subsection{Cryptographic provenance for data artifacts}
\label{sec:background:provenance}

The defense in \Cref{sec:defense} is an application of a known
design family to a new substrate. Four prior systems set the
pattern:

\paragraph{Software-artifact provenance.}
sigstore~\cite{newman2022sigstore} pioneered the model of attaching
a signature to a published software artifact, distributing public
keys via short-lived certificates issued by a transparency-logged
identity oracle. The defining property is that any artifact
reaching a consumer carries enough metadata to verify it without
out-of-band trust establishment.

\paragraph{Supply-chain attestations.}
in-toto~\cite{intoto} extended the sigstore-shaped model to
multi-step pipelines, attaching attestations to each pipeline step
so the consumer can verify not only the final artifact but the
sequence of transformations that produced it. SLSA (Supply-chain
Levels for Software Artifacts)~\cite{slsa} provides a layered
taxonomy of guarantees within this design.

\paragraph{Media provenance.}
The Coalition for Content Provenance and Authenticity~\cite{c2pa}
applies the same family of ideas to images, video, and audio,
binding a content artifact to its capture device, editing
history, and identity claims via a chain of cryptographic
manifests.

\paragraph{Tool-schema provenance.}
SchemaPin~\cite{schemapin-repo} --- a sister project from the
authors --- applies signed-payload provenance to JSON schemas
for tool calls in Model Context Protocol (MCP) deployments, so
that an MCP client can verify a tool's schema before invocation.

VectorPin fills the corresponding gap at the embedding layer.
The wire format is simpler than any of the four prior systems
above because the artifact being attested --- a pinned embedding
plus its source text reference --- is much smaller and has a
fixed structure. The contribution is the canonical byte form for
floating-point arrays, the cross-language test-vector discipline,
and the application of this design family to a substrate that
currently has no provenance story at all. We discuss the
positioning in more detail in \Cref{sec:related}.

\section{Threat Model}
\label{sec:threat_model}

The strongest reviewer objection to a paper of this shape is the
question: \emph{if the attacker can read the raw vectors,
why bother with steganography? Why not exfiltrate the source corpus
directly?} Taking that question seriously is the difference between
an empirically defensible paper and one that overclaims.
We answer it by separating three adversary tiers explicitly,
naming which tier each empirical result actually supports, and
naming the tier we leave out of scope.

\subsection{Three adversary tiers}
\label{sec:threat_model:tiers}

\begin{table}[t]
\centering
\small
\caption{Adversary capability matrix. Threat model A is the
focus of this paper's empirical results. B is supplementary. C is
explicitly out of scope.}
\label{tab:threat_model}
\begin{tabularx}{\linewidth}{Xccc}
\toprule
Capability & A: Insider Backup & B: Compromised DB Creds & C: Query-Only \\
\midrule
Read source documents      & \(\times\) & \(\times\) & \(\times\) \\
Write to ingestion pipeline& \(\checkmark\) & \(\times\) & \(\times\) \\
Read raw vectors           & \(\checkmark\) & \(\checkmark\) & \(\times\) \\
Issue similarity queries   & \(\checkmark\) & \(\checkmark\) & \(\checkmark\) \\
DLP scans source documents & \(\checkmark\) & n/a & n/a \\
DLP scans vector store     & \(\times\) & \(\times\) & \(\times\) \\
\bottomrule
\end{tabularx}
\end{table}

\paragraph{Threat model A: Insider with backup access (primary).}
A privileged engineer in an organization that uses RAG for internal
knowledge retrieval. The engineer can ingest documents into the
pipeline (the legitimate authoring path) and has read access to
vector-store backups --- a reasonable approximation of an SRE,
ML platform engineer, or anyone with the credentials needed to operate
the production system. The engineer cannot exfiltrate raw documents
because those pass through a DLP product, are watermarked, or
are stored in a data room the engineer does not have direct access to.

The engineer's goal is to smuggle sensitive source content out via
the vector backup. They write metadata into the embedding pipeline
at ingestion time and use that metadata to invert the obfuscation
once the backup is exfiltrated. \emph{All capacity, fidelity, and
detection metrics in \Cref{sec:evaluation} are reported under
this model.}

\paragraph{Threat model B: Compromised vector-DB credentials
(secondary).}
An external attacker who has stolen vector-database credentials ---
e.g., an exposed API key, a leaked service principal --- but does
not have access to the source documents because those have been
deleted, stored elsewhere, or encrypted at rest with keys the
attacker does not hold. The attacker cannot write to the ingestion
pipeline; they can only read existing vectors and issue queries.

This model is weaker than A in attack surface (no opportunity to
plant attestation metadata at ingestion) but is real, because
disclosure of vector-DB credentials does happen. We discuss it in
\Cref{sec:discussion} but do not produce empirical numbers under it.

\paragraph{Threat model C: Query-only (out of scope).}
An attacker with similarity-search query access only --- no ability
to read raw vectors. Any payload recovery happens through
rank-information channels, which are bandwidth-limited
relative to direct vector access by orders of magnitude. We mention
this model for completeness but explicitly do not validate the
techniques in \Cref{sec:techniques} against it.

\subsection{Why steganography rather than direct dump?}
\label{sec:threat_model:why_stego}

Under threat model A, a defender will reasonably ask why the insider
does not just \texttt{qdrant export} the entire collection. Three
reasons make steganography preferable to direct extraction:

\paragraph{Egress shaping.} A direct dump is a single large
transfer that trips standard egress monitoring (volume thresholds,
rare-destination heuristics, sudden burst from a service account).
Steganographic payloads piggyback on routine query and write traffic
that is already on the wire and already classified as benign.

\paragraph{Targeted recovery.} The insider often wants specific
high-value documents (a particular financial report, a specific
patient file), not the corpus. Steganographic markers let the
attacker pinpoint the right vectors without first downloading
gigabytes and then sifting through them.

\paragraph{Plausible deniability.} A vector store full of
innocuous-looking embeddings reads as ``business as usual'' to a
forensic investigation. A direct dump leaves an unambiguous
fingerprint: a service account that read every vector in a
collection in a span the attacker chose. The latter is far easier
for an incident-response team to triage.

These reasons are not unique to vector stores --- the same
arguments apply to DNS exfiltration, HTTPS-layer covert channels,
and steganography in image attachments. We restate them here only
because the reviewer-question rate on this paper category demands
it.

\subsection{What VectorPin assumes about the adversary}
\label{sec:threat_model:vectorpin_assumptions}

The defense in \Cref{sec:defense} is designed against an adversary
who can:

\begin{itemize}[leftmargin=*]
\item Modify vectors after they are produced (via a poisoned
  ingestion pipeline, a compromised vector DB, or backup-level
  access).
\item Observe the public verification key, but not the private
  signing key.
\item Replay or selectively delete pin attestations.
\end{itemize}

It does \emph{not} defend against:

\begin{itemize}[leftmargin=*]
\item An attacker with the private signing key. Key custody is the
  user's responsibility.
\item An attacker who modifies source documents \emph{before}
  embedding. Upstream content integrity controls remain necessary.
\item An attacker who uses a legitimate signing key to attest a
  malicious vector at ingestion time. Upstream input validation
  remains necessary.
\end{itemize}

We restate these limits in \Cref{sec:defense:not_caught} after
introducing the protocol so the reader can match each non-claim to
the relevant verification step.

\section{Steganographic Techniques}
\label{sec:techniques}

We separate the techniques into two architecturally distinct
families. \emph{Perturbation attacks} (\Cref{sec:tech:noise}
through \Cref{sec:tech:combined}) operate on a single embedding
\(v \in \mathbb{R}^d\) produced by a model \(M\) and produce a
perturbed embedding \(v' \in \mathbb{R}^d\) that an attacker writes
to the vector store in place of \(v\); the hidden payload is encoded
in the difference \(v' - v\). \emph{Architectural smuggling} attacks
(\Cref{sec:tech:fragmentation}) do not modify a clean embedding at
all --- they exploit the architectural choice of which embedding
model produces which fragment of the source content. Only the first
family is statistically detectable in principle; the second
requires architectural defenses (model attestation) rather than
distributional ones. The empirical evaluation in
\Cref{sec:evaluation} reports detector AUCs only for the
perturbation family; fragmentation is addressed by the
\texttt{model} field of the VectorPin attestation rather than by
statistical detection.

\Cref{tab:techniques_overview} summarizes the families, their
parameters, recovery-metadata requirements, and a payload-bit proxy
that gives an upper bound on the encodable channel size per vector.
The proxy is not a Shannon-capacity proof; it is an operational
bound on \emph{how much} an attacker could in principle hide given
the parameter freedom each technique offers.

\begin{table}[t]
\centering
\small
\caption{Steganographic techniques studied, grouped by architectural
family. The \emph{recovery metadata} column lists what the attacker
must store at ingestion time to invert the transformation later;
this metadata is what VectorPin's signed attestation prevents the
attacker from substituting. The \emph{payload-bit proxy} column is
an upper bound on encodable channel size per vector at typical
operating parameters. ``low'' means under 100 bits, ``medium''
100-1000, ``high'' over 1000.}
\label{tab:techniques_overview}
\begin{tabularx}{\linewidth}{llXll}
\toprule
Family & Technique & Recovery metadata & Invertible? & Payload bits \\
\midrule
\multirow{5}{*}{Perturbation} & Noise injection & \(\{n_i\}\) (samples or seed) & w/ samples or seed & high \\
                              & Rotation        & \(R\) (matrix or seed)        & exact              & medium-high \\
                              & Scaling         & \(\{s_i\}\)                   & exact              & low \\
                              & Offset          & \(o\)                         & exact              & low (single channel) \\
                              & Combined        & all of the above              & w/ all             & high \\
\midrule
Architectural & Fragmentation & dimension partition, model ids & exact, w/ map & medium (in partitioning) \\
\bottomrule
\end{tabularx}
\end{table}

\subsection{Noise injection}
\label{sec:tech:noise}

The simplest technique. For each embedding, sample
\(n \sim \mathcal{N}(0, \sigma^2 I_d)\) and emit
\(v' = v + n\). The attacker stores \(\{n_i\}\) as recovery
metadata so that, given the obfuscated batch, they can
recover \(v_i = v'_i - n_i\). The hidden payload is encoded
in the noise samples themselves --- e.g., the bit string
\(b\) is mapped to a deterministic pseudorandom \(n(b)\) so the
recipient who knows \(b\) (or the seed that produced it) can
re-derive the noise and decode.

The capacity-detectability trade is controlled by \(\sigma\).
For a unit-norm embedding, \(\sigma\) acts as a per-dimension
amplitude budget: small \(\sigma\) preserves cosine similarity
but is hard to encode meaningful payload into; large \(\sigma\)
moves the embedding off the unit sphere and shows up as a
distributional shift visible to per-dimension variance tests.
We measure three values
(\(\sigma \in \{0.005, 0.01, 0.05\}\)) plus the obfuscator's
library default in \Cref{sec:evaluation:fidelity}, where the
operating point at \(\sigma = 0.005\) preserves \(\cos = 0.964\)
but is already detectable by One-Class SVM at AUC \(0.796\).

\subsection{Rotation}
\label{sec:tech:rotation}

For \(R \in O(d)\) (an orthogonal matrix), \(v' = R v\) leaves
the magnitude \(\|v'\| = \|v\|\) and the pairwise inner products
\(\langle R u, R v \rangle = \langle u, v \rangle\) unchanged.
Cosine similarity is therefore invariant under any consistent
rotation applied to all stored vectors and queries --- which is
the hardest case for a defender, because no per-vector statistic
distinguishes the obfuscated batch from a clean one as long as
the same \(R\) is applied uniformly.

In our implementation \(R\) is constructed by composing several
random Givens rotations of bounded angle:
\[
R = \prod_{(i,j,\theta) \in \Pi} G_{i,j}(\theta), \quad
\theta \in [-\theta_{\max}, +\theta_{\max}]
\]
The hidden payload is encoded in the choice of \((i, j, \theta)\)
tuples; the attacker stores \(R\) (or the seed that generated it)
as recovery metadata.

Larger \(\theta_{\max}\) and larger \(|\Pi|\) move the rotated
vector farther from its original (we measure cosine drops to
0.68 at \(\theta_{\max} = \pi/2\), \(|\Pi| = 200\) ---
\Cref{sec:evaluation:rotation_sweep}), but the rotated batch
remains distributionally indistinguishable to the detectors we
evaluate because the transformation is orthogonal. The limiting
factors on the attacker's parameter choice are therefore
retrieval compatibility, payload encoding efficiency, and whether
a downstream verifier checks against the original vector ---
\emph{not} anomaly-detection visibility.

\subsection{Scaling}
\label{sec:tech:scaling}

Per-vector magnitude rescaling: \(v'_i = s_i \cdot v_i\) for
scalars \(s_i\). Direction is preserved (so cosine similarity
to other unscaled vectors changes only as \(\cos\theta\) where
\(\theta\) is the rescaling-induced angle, which is zero), but
norm-based detectors see the per-vector scale shift directly.
The attacker stores \(\{s_i\}\) as recovery metadata.

In practice this technique is mostly useful when combined with
others: alone, it produces no information channel except through
the choice of \(s_i\) values, which are visible in
\(\|v'_i\|/\|v_i\|\) if the defender ever sees both. Our
evaluation reports it as a control --- AUC is essentially random
at the library default (\Cref{tab:detection}) because the
default scale shift is too small to register against the
intrinsic norm variance of the corpus.

\subsection{Offset}
\label{sec:tech:offset}

A constant additive shift: \(v' = v + o\) for a fixed
\(o \in \mathbb{R}^d\). This is the most destructive technique
in our evaluation because the shared offset shows up immediately
in pairwise statistics: every pair of obfuscated vectors has
the same offset added, so the pairwise differences
\((v'_i - v'_j) = (v_i - v_j)\) are unchanged but the centroid
of the batch shifts by \(o\). Defenders fit on a clean baseline
see this immediately as a shift in the empirical mean.

The technique is included in our study not because it is
practical for real attackers --- it is not --- but because it
demonstrates the upper bound of our detection results: at
\(o\) magnitude \(0.05\), Isolation Forest and One-Class SVM
both reach AUC \(1.000\) and TPR@1\%FPR \(= 1.000\). It is
the easy case for the defender.

\subsection{Cross-model fragmentation}
\label{sec:tech:fragmentation}

Rather than perturbing a single embedding, the attacker splits
the source content across multiple embedding models
\(\{M_1, M_2, \ldots, M_k\}\) and stores the resulting
fragments in separate vector stores or under separate
collection names. Each individual store sees only a fragment;
recovery requires joining the fragments under the original
partition map. The recovery metadata is the partition: which
fragment lives where, under which model, in which order.

This is a different kind of attack from the perturbation
techniques --- there is no in-place modification of a clean
embedding, so no statistical anomaly is induced. The defense in
\Cref{sec:defense} addresses this case via the
\texttt{model} field of the pin attestation: a verifier
checking that all retrieved fragments come from the expected
model and collection raises \texttt{MODEL\_MISMATCH} as soon
as fragments from a foreign model appear. This is an architectural
defense rather than a statistical one, and it generalizes to
any cross-model exfiltration variant the attacker might design.

\subsection{Combined techniques}
\label{sec:tech:combined}

Sequential composition of the above. The implementation we use
applies the obfuscator's full pipeline as
noise \(\to\) rotation \(\to\) scaling \(\to\) offset, with
each stage receiving the output of the previous.

The empirical interest of the combined technique is that
\emph{it does not gain stealth from composition}. Detector AUC
on the combined technique tracks the most-detectable component
in the chain, which in our defaults is offset. The intuition is
that off-the-shelf detectors fit a single density model over
the embedding distribution; whichever sub-technique pushes a
sample furthest off-density dominates the score. We discuss
this further in \Cref{sec:appendix:negative}.

\subsection{Capacity bounds}
\label{sec:tech:capacity}

A natural question: how many bits of payload does each
technique carry per embedding? The honest answer is that
capacity is hard to express as a fixed number for two reasons.

First, the channel is parameter-dependent. For noise injection
at \(\sigma = 0.005\) the informational capacity per dimension
is bounded by Shannon-Hartley with a noise floor that depends
on the precision of the recipient's recovery; at \(\sigma = 0.05\)
that capacity rises but the technique becomes detectable.
Second, the payload is constrained by the recovery-metadata
budget: the attacker must store enough metadata at ingestion
to invert later, and the metadata channel itself is observable.
A fully self-contained capacity proof would need to model the
metadata channel as part of the cover signal, which we do not.

For this reason we report fidelity, detectability, and recovery
cosine in \Cref{sec:evaluation} rather than a single
bits-per-dimension number. The \(\sigma\)-sweep in
\Cref{tab:detection} traces out the operating curve, and a
reader interested in capacity at a given operating point can
read it off the curve directly. For the rotation channel
specifically we additionally derive a closed-form per-vector
capacity \(\lfloor d/2 \rfloor \cdot b\) bits with a working
encoder/decoder, separate the capacity ceiling from the
retrieval-preserving operating point, and characterise how the
detector AUC depends on the clean-corpus distribution
(\Cref{sec:evaluation:capacity}). A fully general rate-distortion
analysis with paraphrased-query retrieval as the distortion metric
remains future work (\Cref{sec:limitations}); the rotation case is
done here.

\subsection{Recovery metadata: what the attacker must keep}
\label{sec:tech:recovery}

\Cref{tab:techniques_overview} summarizes the recovery
metadata for each technique. Note the key property: \emph{every
technique requires post-embedding modification of the stored
vector.} Noise injection adds \(n\); rotation applies \(R\);
scaling multiplies by \(s\); offset adds \(o\); fragmentation
substitutes one embedding model's output for another. None of
the techniques produces a vector that the issuer's model
\(M\) would have produced for the source content.

This is the structural property VectorPin exploits in
\Cref{sec:defense}. The \texttt{vec\_hash} field of the signed
attestation commits to the model's actual output. Any
post-embedding modification --- regardless of which technique
in this section, regardless of the recovery metadata the
attacker stores, regardless of how cleverly the perturbation is
designed to evade statistical detectors --- changes
\texttt{vec\_hash} and triggers \texttt{VECTOR\_TAMPERED} on
verification. The attacker's only remaining options are to
forge the signature (which requires the private signing key)
or to attest the malicious vector at ingestion time (which
requires control of the ingestion pipeline, which is a
strictly stronger threat model than the one studied here).

\section{Empirical Evaluation}
\label{sec:evaluation}

This section reports the empirical study at the heart of the paper.
We measure five things per technique \(\times\) operating point:
fidelity (does the embedding still represent the source content),
distributional detectability (does an off-the-shelf anomaly detector
flag it), quantization resilience (does the perturbation survive the
int8 round-trip that production vector stores apply), retrieval
utility (does a legitimate user still get the right document back),
and performance (per-batch wall clock).

\subsection{Setup}
\label{sec:evaluation:setup}

\paragraph{Embedding model.} \texttt{text-embedding-3-large}
(OpenAI), \(3072\)-dimensional output, accessed via the official API.
We re-embedded the corpus once at the start of the run and committed
the resulting matrix as a NumPy file (\texttt{embeddings\_baseline
.npy}) so subsequent re-evaluations are deterministic.

\paragraph{Corpus.} A synthetic 68-chunk corpus drawn from the
\texttt{sample\_docs/} directory of the
VectorSmuggle repository~\cite{vectorsmuggle-repo}. The corpus
contains synthetic-PII documents (employee records, payroll data,
financial reports, API documentation) deliberately constructed to
trigger any reasonable sensitive-data classifier. The synthetic
nature is a feature, not a limitation: it keeps the experiment
reproducible without releasing real personal data.

\paragraph{Reproducibility.} All randomness uses fixed seeds
(\texttt{seed=42}). The empirical-study driver
(\fpath{scripts/empirical_study.py}) and the preprint-extension
driver (\fpath{scripts/preprint_extensions.py}) are checked into
the public repository. \Cref{sec:appendix:reproducibility} lists
the exact commands.

\paragraph{Configurations.} We sweep four noise configurations
(\texttt{noise\_low}~$\sigma=0.005$, \texttt{noise\_med}~$\sigma=0.01$,
\texttt{noise\_high}~$\sigma=0.05$, and \texttt{default} which uses
the obfuscator's library defaults) and seven techniques per
configuration: \emph{noise}, \emph{rotation}, \emph{scaling},
\emph{offset}, \emph{noise+rotation}, \emph{noise+rotation+scaling},
and \emph{all} (the obfuscator's combined pipeline). This yields
\(28\) measurement points for fidelity and recovery.

For detection and quantization analysis we report the six
single-technique configurations (with the noise variants reported at
\(\sigma \in \{0.005, 0.01, 0.05\}\) and rotation/scaling/offset at
their library defaults), against detectors fit on a held-out clean
batch.

\paragraph{Detector training.} The 68-chunk embedding matrix was
randomly split (RNG seed 42) into 27 training vectors and 41 test
vectors. Both Isolation Forest and One-Class SVM were fit on the
27 clean training vectors only. The 41 test vectors served as the
clean-side scoring set for ROC computation; obfuscated versions of
those same 41 test vectors served as the attack-side scoring set.
This splits-on-honest-data design ensures the detector never sees
a perturbed sample at fit time, matching the realistic operator
flow of training the detector during a known-clean baseline window.

\subsection{Fidelity: how much does the embedding move?}
\label{sec:evaluation:fidelity}

\Cref{tab:fidelity} reports per-vector cosine similarity between the
original and obfuscated embedding, the correlation of the pairwise
similarity matrix before and after obfuscation (an operationally
meaningful measure --- if pairs that were close before are still
close after, retrieval downstream is largely preserved), and end-to-end
recovery cosine for the techniques whose
metadata permits inversion.

\begin{table}[t]
\centering
\small
\caption{Fidelity metrics. \texttt{cos(orig,obf)} is the per-vector
cosine similarity between the clean and obfuscated embedding;
\texttt{pair-corr} is the Pearson correlation between the upper
triangles of the pairwise cosine similarity matrices before and
after obfuscation; \texttt{rec-cos} is recovery cosine after
applying inverse transformations using attestation metadata (--- =
not invertible).}
\label{tab:fidelity}
\begin{tabular}{lrrr}
\toprule
Technique             & cos(orig,obf) & pair-corr & rec-cos \\
\midrule
rotation @ 0.1 rad    & 1.000 & 1.000 & 1.000 \\
scaling @ 0.95        & 1.000 & 1.000 & 1.000 \\
noise @ \(\sigma{=}0.005\)  & 0.964 & 0.998 & --- \\
noise @ \(\sigma{=}0.01\)   & 0.874 & 0.983 & --- \\
offset @ 0.05         & 0.527 & 0.735 & --- \\
noise @ \(\sigma{=}0.05\)   & 0.336 & 0.571 & --- \\
\midrule
noise+rotation (med)  & 0.875 & 0.982 & 0.875 \\
noise+rot+scale (med) & 0.875 & 0.982 & 0.875 \\
all (med)             & 0.490 & 0.674 & 0.490 \\
\bottomrule
\end{tabular}
\end{table}

\begin{figure}[t]
\centering
\includegraphics[width=\linewidth]{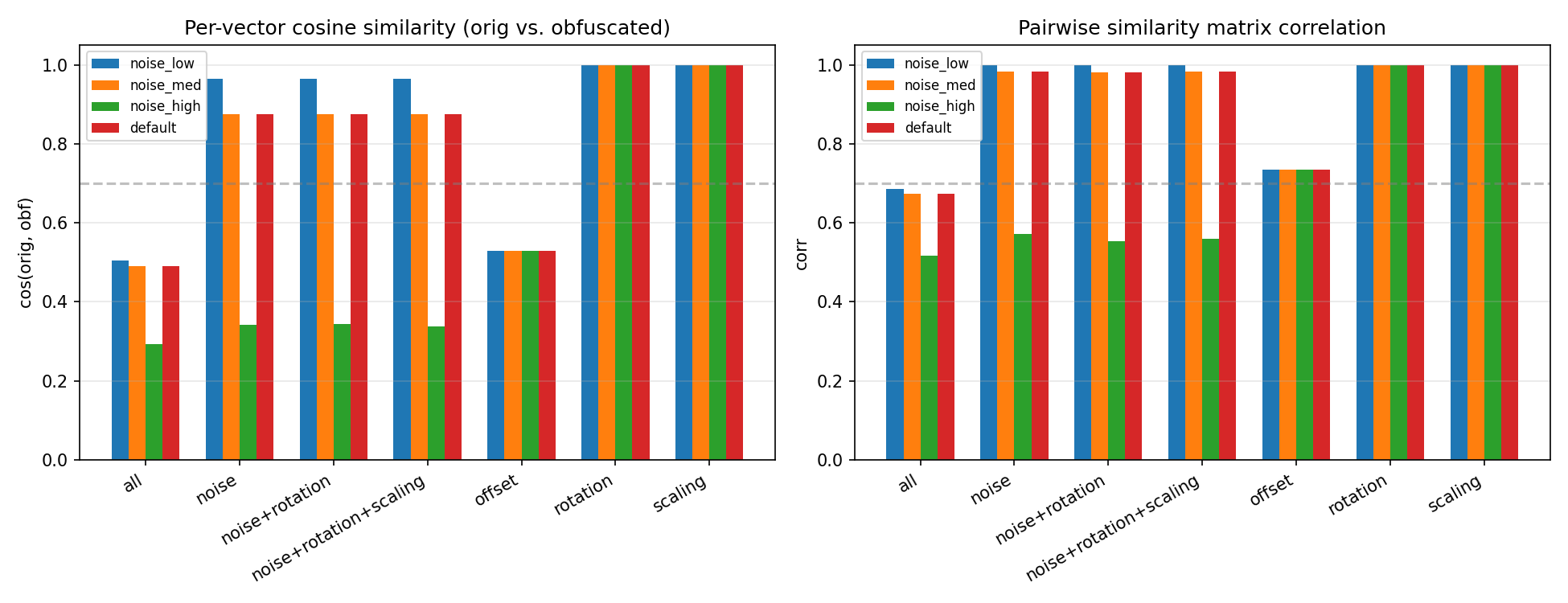}
\caption{Per-vector cosine similarity (left) and pairwise similarity
matrix correlation (right) across techniques and noise levels.
The dashed line at 0.7 marks a rough operational threshold: below it
retrieval quality begins to visibly degrade.}
\label{fig:fidelity}
\end{figure}

\Cref{fig:fidelity} visualizes the same data across all four noise
configurations.

Two observations matter for the rest of the paper:

\begin{enumerate}[leftmargin=*]
\item \textbf{Rotation and scaling at the obfuscator's default
parameters are nearly invisible to fidelity.} Both produce
\(\cos = 1.000\) and \texttt{pair-corr} \(= 1.000\). This
is consistent with theory: small-angle Givens rotations and
small magnitude scaling preserve direction, and cosine similarity
is direction-only. \emph{Visibility to fidelity} is not the same
as \emph{visibility to a defender}, which we measure next.
\item \textbf{Offset at the default range is destructive even before
detection enters the picture.} A common additive offset across all
test vectors moves them along a shared direction, dropping
\(\cos\) to \(0.527\). This is detectable simply as ``the
embeddings don't look right anymore'' before any statistical
detector is involved.
\end{enumerate}

\subsection{Detection: ROC against off-the-shelf anomaly detectors}
\label{sec:evaluation:detection}

We trained two detectors~\cite{pedregosa2011scikit} on the 27-vector
clean-training split:

\begin{itemize}[leftmargin=*]
\item \textbf{Isolation Forest}~\cite{isolationforest} (sklearn
defaults: \(n\_\text{estimators} = 200\),
\(\text{contamination} = 0.05\), random state 42).
\item \textbf{One-Class SVM}~\cite{scholkopf2001ocsvm} with RBF
kernel and \(\nu = 0.05\).
\end{itemize}

For each technique \(\times\) operating point we scored the 41
clean test vectors and the 41 obfuscated versions of those test
vectors, treating the labels as 0 (clean) and 1 (obfuscated) and
computing ROC-AUC, F1 at sklearn's default threshold, and
TPR at 1\% and 5\% false-positive rate cuts.

\Cref{tab:detection} reports the headline numbers.

\begin{table}[t]
\centering
\small
\caption{Detection metrics for distribution-shifting perturbations.
AUC \(\geq 0.85\) on at least one detector at every tested
non-rotation operating point that meaningfully moves embedding
direction. The rotation row is included as a baseline reference;
orthogonal rotation has no detector signal at any parameter setting
and is handled separately in \Cref{sec:evaluation:rotation_sweep}.
TPR @ 1\% FPR shows the rate at which a defender willing to flag
1\% of legitimate traffic catches the attack.}
\label{tab:detection}
\begin{tabular}{lrrrr}
\toprule
Technique  & IF AUC & OCSVM AUC & IF TPR@1\%FPR & IF TPR@5\%FPR \\
\midrule
rotation @ 0.1 rad     & 0.498 & 0.501 & 0.000 & 0.024 \\
scaling @ 0.95         & 0.404 & 0.289 & 0.024 & 0.073 \\
noise @ \(\sigma{=}0.005\) & 0.638 & 0.796 & 0.098 & 0.317 \\
noise @ \(\sigma{=}0.01\)  & 0.864 & 1.000 & 0.390 & 0.659 \\
offset @ 0.05          & 1.000 & 1.000 & 1.000 & 1.000 \\
noise @ \(\sigma{=}0.05\)  & 1.000 & 1.000 & 1.000 & 1.000 \\
\bottomrule
\end{tabular}
\end{table}

\begin{figure}[t]
\centering
\includegraphics[width=\linewidth]{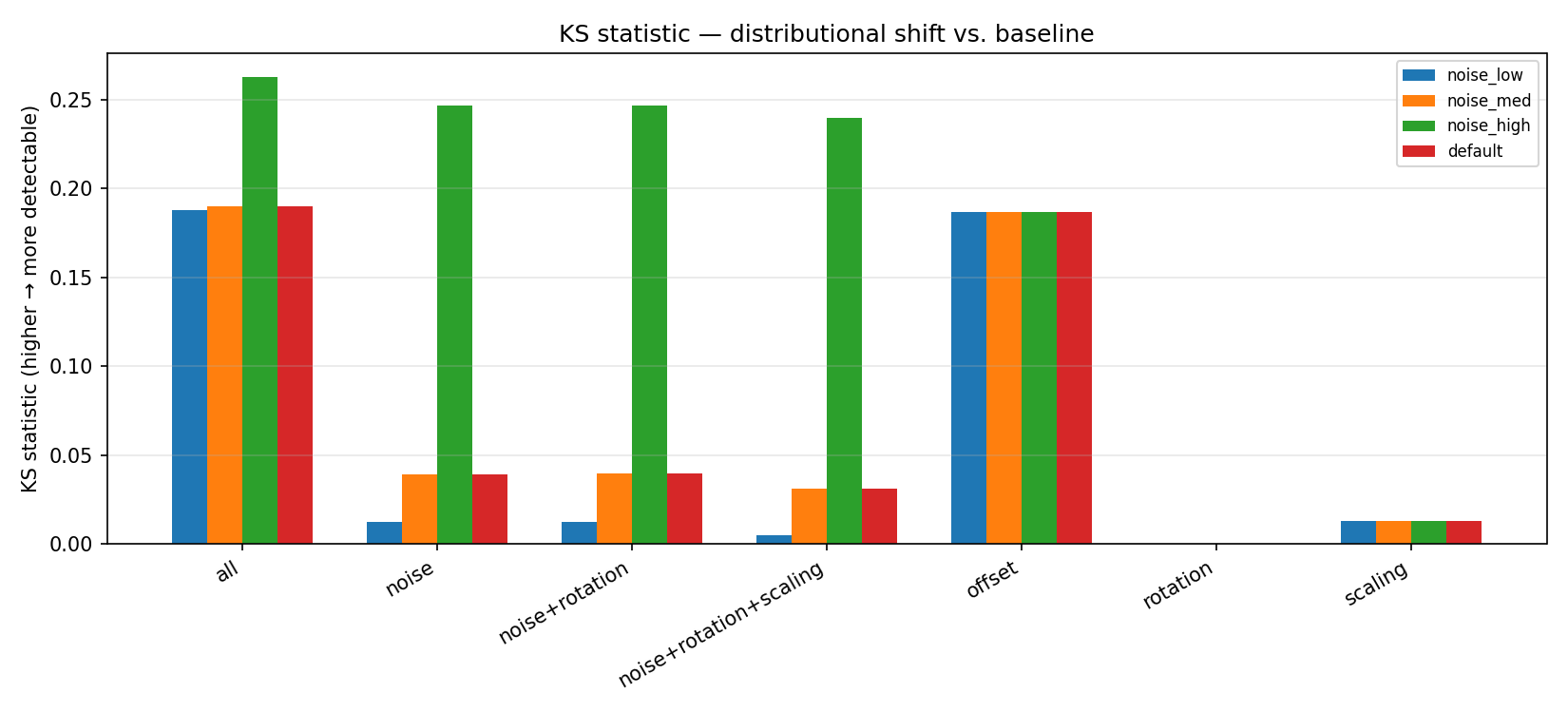}
\caption{Kolmogorov-Smirnov statistic between clean and obfuscated
embedding-component distributions. Higher KS \(=\) more visible
distributional shift.}
\label{fig:detection}
\end{figure}

\Cref{fig:detection} plots the KS-test statistic between clean and
obfuscated embedding distributions across configurations as a
distribution-shift companion view.

\paragraph{Headline finding.} The default rotation row in
\Cref{tab:detection} is the configuration that defeats both
detectors at \(\cos = 1.000\) and AUC \(\approx 0.5\); it is not
itself a high-capacity operating point. The matching question of
how much payload rotation can carry as the rotation parameters
grow is the subject of \Cref{sec:evaluation:capacity}, with the
fidelity-vs-capacity-vs-detectability trade-off characterised
there. The narrower observation here is: at the parameters used by
the obfuscator's library defaults, rotation produces no detector
signal in this corpus. Scaling at the library default produces AUC
\emph{below} \(0.5\) on both detectors, which is a degenerate
result rather than evasion: the perturbation is so small that the
detector finds the obfuscated batch \emph{more} uniform than the
clean training batch in this corpus, so it preferentially flags
clean vectors.

\paragraph{The capacity-detection frontier.} The honest reading of
\Cref{tab:detection} is that the attacker faces a sharp trade.
At \(\sigma = 0.005\) noise, the technique still preserves
\(\cos = 0.964\) but One-Class SVM already reaches AUC \(0.796\) ---
detectable at \(31.7\%\) TPR for a defender willing to false-positive
at \(5\%\). At \(\sigma = 0.01\) noise, AUC reaches \(1.000\) on
OCSVM. The window between ``too small to hide anything'' and
``trivially detectable'' is a narrow band in this corpus.

\subsection{Quantization resilience}
\label{sec:evaluation:quantization}

Production vector stores apply quantization to reduce memory and
disk footprint. Qdrant's default scalar quantization stores each
dimension as int8, scaled by per-dimension global min/max bounds
across the indexed corpus. We modeled this round trip explicitly
and re-measured cosine similarity post-quantization.

\Cref{tab:quantization} reports the comparison. Per-dimension int8
preserved cosine to three decimal places across every technique we
tested. The intuition is that quantization noise at int8 resolution
(\(\approx 1/255\) per dimension after scaling) is smaller than
any of the perturbations we apply, so the technique-induced shift
dominates the quantization shift and survives the round trip.

\begin{table}[t]
\centering
\small
\caption{Cosine similarity to original embedding before and after
per-dimension int8 quantization (Qdrant-default model). The
column \texttt{cos-drop} is \texttt{cos(pre)} \(-\) \texttt{cos(post)}.}
\label{tab:quantization}
\begin{tabular}{lrrr}
\toprule
Technique  & cos(pre-q) & cos(post-q) & cos-drop \\
\midrule
rotation @ 0.1 rad        & 1.000 & 1.000 & 0.000 \\
scaling @ 0.95            & 1.000 & 1.000 & 0.000 \\
noise @ \(\sigma{=}0.005\)& 0.964 & 0.964 & 0.000 \\
noise @ \(\sigma{=}0.01\) & 0.874 & 0.874 & 0.000 \\
offset @ 0.05             & 0.527 & 0.527 & 0.000 \\
noise @ \(\sigma{=}0.05\) & 0.336 & 0.336 & 0.000 \\
\bottomrule
\end{tabular}
\end{table}

This is good news for the attacker and bad news for the defender
who hoped quantization would silently destroy hidden payloads.
Per-vector quantization (used in some FAISS PQ variants but not the
default for general-purpose vector stores) is more aggressive
because each vector receives its own scale-and-shift, which can move
direction; we did not evaluate it here. Product quantization is
evaluated separately in the cross-backend round-trip
(\Cref{sec:evaluation:cross_backend}) at FAISS IVF-PQ \(m{=}8\) and
\(m{=}16\); it is the one configuration where the attacker's bit
channel is materially narrowed (\texttt{recover\_cos} drops to
\(0.76\)--\(0.88\)), as a side effect of memory optimisation rather
than an intentional defense. Binary embeddings remain future work
(\Cref{sec:limitations}).

\subsection{Retrieval utility: motivation}
\label{sec:evaluation:retrieval}

A legitimate user of a RAG system queries with text, the system
embeds the query, and similarity search returns the nearest
documents. If obfuscation breaks this, downstream RAG quality
monitoring will surface the attack regardless of any explicit
defense.

The natural test --- using each test embedding as a self-query
against the obfuscated corpus --- is trivial because the same
perturbation is applied to query and corpus, so the nearest
neighbor remains the source vector and recall is \(1.0\) by
construction. The faithful test, presented in
\Cref{sec:evaluation:paraphrased}, embeds an LLM-generated
\emph{paraphrase} of the source document as the query and checks
whether the source document is still in the top-\(k\) when the
corpus has been obfuscated but the query has not.

\subsection{Performance}
\label{sec:evaluation:performance}

\begin{figure}[t]
\centering
\includegraphics[width=\linewidth]{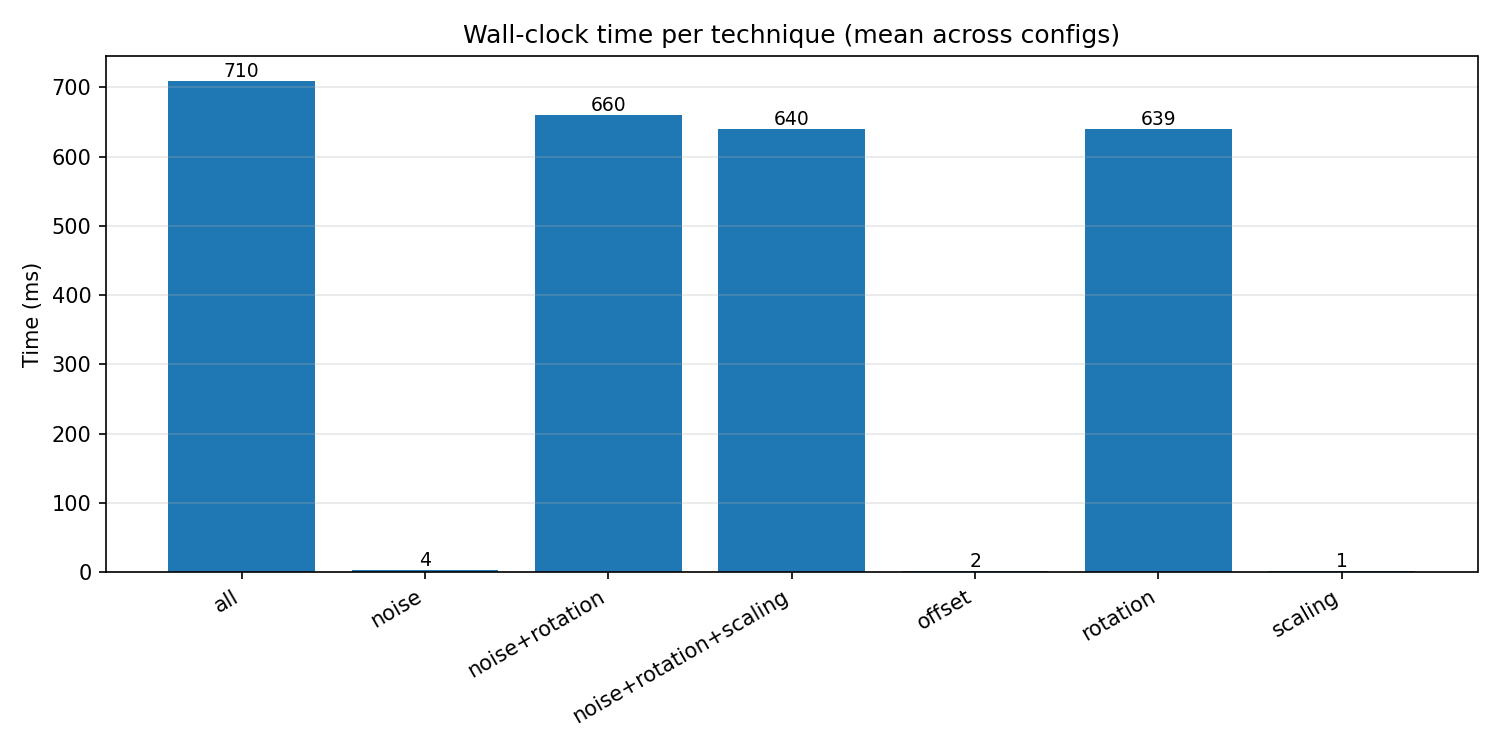}
\caption{Per-technique wall-clock per batch (mean across noise
configurations). Bars represent CPU-time milliseconds for the
full pipeline applied to the 41-vector test split.}
\label{fig:performance}
\end{figure}

\Cref{fig:performance} shows mean wall-clock per technique
across the four noise configurations.

The dominant cost is in the rotation step (random Givens rotations
on a \(3072\)-dim vector); all other steps run in tens of
milliseconds per batch. From a defender's perspective this is
relevant because it confirms that \emph{neither attack nor defense
is throttled by compute}: an attacker can apply these techniques
at ingestion line rate, and a defender can run a real-time detector
in the same loop without changing the system's latency profile.

\subsection{Rotation parameter sweep}
\label{sec:evaluation:rotation_sweep}

The detection results in \Cref{sec:evaluation:detection} reported
rotation at the obfuscator's library default, which is conservative
on both axes (small \(\theta_{\max}\), few Givens rotations). To
test whether more aggressive rotation parameters would let the
detector eventually catch up, we ran a parameter sweep over
\(\theta_{\max} \in \{0.01, 0.05, 0.1, 0.5, 1.0, \pi/4, \pi/2\}\)
and \(\text{num\_rotations} \in \{1, 5, 10, 50, 200\}\) ---
35 operating points in total.

The result is decisive for the detectors evaluated here, on this
clean distribution. Across the entire sweep neither detector
achieves AUC above \(0.53\). Cosine to original drops as parameters
increase (down to \(0.68\) at \(\theta_{\max} = \pi/2\),
\(\text{num\_rotations} = 200\)), but the resulting embeddings
remain indistinguishable from the clean distribution to both
Isolation Forest and One-Class SVM. The structural argument is that
rotation by an orthogonal matrix preserves all pairwise inner
products, all per-dimension marginal moments, and all norm-based
statistics --- the population-level feature surface that
distribution-based anomaly detectors fit a density model on. With
this caveat: the sweep here uses 300-vector unit-Gaussian splits, so
the trained density model has no concentrated manifold to fit. On a
real production embedding distribution the local-manifold geometry
is tighter and a rotation can move samples off the trained-on
manifold, opening a detector signal at high rotation counts; see
\Cref{sec:evaluation:capacity}, \Cref{tab:capacity_real} for the
analogous run on real \fpath{text-embedding-3-large} corpus vectors,
where the rotation-AUC behaviour at high \(K\) is markedly different.
The small-angle / few-rotation regime that drives the
\(\cos \approx 1\) operating point in \Cref{sec:evaluation:detection}
remains undetectable on real corpus vectors at the parameters used
there.

\begin{table}[t]
\centering
\small
\caption{Rotation sweep selected operating points, full table in the
supplementary results bundle. Entries are mean per-vector cosine
similarity to original, Isolation Forest AUC, and One-Class SVM AUC,
all computed against detectors trained on a held-out 300-vector
clean batch with the perturbation applied to a separate 300-vector
target batch in 384-dimensional space. Across all 35 sweep points
neither detector exceeded \(0.53\) AUC.}
\label{tab:rotation_sweep}
\begin{tabular}{rrrrr}
\toprule
\(\theta_{\max}\) & num\_rot & cos & IF AUC & OCSVM AUC \\
\midrule
0.01    & 1   & 1.0000 & 0.529 & 0.490 \\
0.01    & 200 & 1.0000 & 0.530 & 0.490 \\
0.10    & 1   & 1.0000 & 0.529 & 0.490 \\
0.10    & 200 & 0.9983 & 0.528 & 0.491 \\
0.50    & 200 & 0.9582 & 0.526 & 0.492 \\
1.00    & 200 & 0.8473 & 0.521 & 0.495 \\
\(\pi/4\) & 200 & 0.9014 & 0.532 & 0.493 \\
\(\pi/2\) & 200 & 0.6804 & 0.494 & 0.498 \\
\bottomrule
\end{tabular}
\end{table}

This finding sharpens the role of rotation in the paper. Rotation
is not merely a ``degenerate'' technique with a bad
capacity-stealth trade; on this synthetic-Gaussian sweep
distribution-based detection has no purchase at any parameter
setting tested, including settings that drop cosine to \(0.68\).
The matching question on real corpus distributions ---
\Cref{sec:evaluation:capacity}, \Cref{tab:capacity_real} --- gives
a more nuanced answer: the small-angle / few-rotation regime
(matching the \(\cos \approx 1\) operating point of
\Cref{sec:evaluation:detection}) stays undetectable, but
random-angle disjoint-Givens rotations at high capacity become
detectable as they move samples off the trained-on local manifold.
A durable defense against rotation should not rely solely on the
obfuscated batch's distribution looking different from clean: in
the small-angle regime there may be no signal at all, while in the
high-capacity regime the signal depends on corpus geometry and
training-set coverage that a defender has to guarantee out-of-band.
\Cref{sec:defense} ships exactly such a corpus-geometry-independent
control via VectorPin's \texttt{vec\_hash} commitment.

\subsection{Empirical payload capacity}
\label{sec:evaluation:capacity}

The rotation sweep in \Cref{sec:evaluation:rotation_sweep} measured
\emph{detectability}; a reviewer-defensible follow-up is to convert
the ``rotation is undetectable'' result into a quantitative attacker
channel. This subsection derives the per-vector capacity in closed
form, ships a working encoder/decoder for the channel, and reports
the detector AUC behaviour as the channel fills --- separately on
the synthetic-Gaussian baseline and on real
\fpath{text-embedding-3-large} corpus vectors. The two regimes
differ, and the difference itself is part of the result.

\paragraph{Closed-form capacity.} Consider a sender and receiver who
share a key that selects \(K\) disjoint dimension pairs
\((i_k, j_k)\) of the embedding (each dimension index used in at
most one pair). The sender encodes \(b\) payload bits per pair as a
quantized angle \(\theta_k \in (-\pi, \pi]\) and applies the
corresponding Givens rotation. Disjoint pairs commute, so the applied
rotations form a block-diagonal orthogonal matrix
\(R = \bigoplus_k G_{i_k j_k}(\theta_k)\). Given the original \(v\)
and the rotated \(R v\), the receiver recovers each angle
independently as
\(\theta_k = \operatorname{atan2}((Rv)_{j_k}, (Rv)_{i_k})
            - \operatorname{atan2}(v_{j_k}, v_{i_k}) \pmod{2\pi}\),
and the per-vector capacity of this channel is
\begin{equation}
C_{\mathrm{disjoint}}(d, b) \;=\; \lfloor d/2 \rfloor \cdot b
\quad \text{bits},
\label{eq:capacity}
\end{equation}
which at \(d = 3072\) (matching \fpath{text-embedding-3-large}) and
\(b = 10\) (a \(2^{10}\)-level angle quantization) is
\(15{,}360\) bits = \(1{,}920\) bytes per vector.

The looser proxy used in the rotation sweep counts an additional
\(\log_2 \binom{d}{2}\) bits per rotation for the choice of
\((i, j)\). That proxy is a strict upper bound on what an idealized
encoding could in principle carry, but it is \emph{not} what a
single-vector decoder can recover: peeling the \((i_k, j_k)\)
sequence out of a rotated vector requires either a shared key (in
which case those bits are not data) or knowledge of the intermediate
states (which the receiver does not have). Disjoint Givens with
shared keyed pairs is the channel an attacker can demonstrate
end-to-end.

\paragraph{Decoder access assumptions.} The recovery formula
requires the receiver to have both \(v\) and \(Rv\). Under our
threat model (\Cref{sec:threat_model:tiers}) this matches the
\emph{attacker-controlled-corpus} case: the accomplice authored or
otherwise possesses the corpus rows being smuggled, so the clean
\(v\) is by construction known at decode time. Three weaker access
modes are worth distinguishing because they would be required to
extend the channel beyond the rows the attacker controls:
\begin{itemize}[leftmargin=*,itemsep=2pt]
\item \emph{Known-cover}: receiver has \(v\) by some out-of-band
means (a sender-side cache of pre-rotation copies, an exfiltrated
log). Capacity matches the demonstration, but the side cache is
itself an exfiltration channel; this case is rarely independently
interesting.
\item \emph{Shared-key differential}: sender and receiver share a
cryptographic key but no \(v\). Recovery from \(Rv\) alone would
require a different protocol --- one that encodes the payload in an
invariant of the rotation that survives the unknown pre-rotation
state. We do not propose or demonstrate such a protocol.
\item \emph{Metadata-assisted}: the attacker stores enough recovery
metadata next to the rotated vector that the receiver can decode.
This relocates the secret onto the metadata channel, and is
interesting only if that channel has the same defensive blind spots
that the rotation channel does; we do not assume it.
\end{itemize}
The empirical results in \Cref{tab:capacity,tab:capacity_real}
all instantiate the attacker-controlled-corpus case: the attacker
authored the corpus rows, the receiver later decodes them back from
the store, and \(v\) is shared as a side effect of the attacker
controlling the ingestion path. Generalising the demonstration to a
threat model where the receiver does \emph{not} have \(v\) would
require an additional protocol-design step that is open work; this
is recorded in \Cref{sec:limitations}.

\paragraph{Empirical decode (synthetic baseline).}
\Cref{tab:capacity} reports a working encoder/decoder for the
disjoint-Givens channel at \(d=3072\), \(b=10\), across payload
sizes from 64 B (\(K=52\)) to the full
\(K_{\max} = \lfloor d/2 \rfloor = 1536\) operating point. Each row
encodes a fresh random payload, applies the corresponding rotation
to a batch of 300 unit-norm clean Gaussian vectors, round-trips the
batch through the listed storage \texttt{dtype} (a real vector DB
rarely stores \texttt{float64}), then decodes vector \([0]\) and
reports the bit error rate. The same Isolation Forest and One-Class
SVM detectors as in \Cref{sec:evaluation:detection} are retrained
per row on a clean held-out 300-vector batch.

Across all 21 operating points the decoded bytes match the input
exactly (BER \(=0\) at every row), including under \texttt{float32}
and \texttt{float16} storage round-trips: the angle-quantization cell
width \(2\pi / 2^b\) exceeds storage roundoff for \(b\) up to
\(\sim 12\). On this synthetic-Gaussian baseline detector AUCs stay
in IF \(\in [0.425, 0.490]\), OCSVM \(\in [0.501, 0.542]\) ---
statistical noise around \(0.5\) --- across the entire sweep,
including at \(K_{\max}\). The synthetic-data result confirms that
the BER claim of the channel is solid; whether the AUC claim
extends to real corpus distributions is a separate question, treated
in the next paragraph.

\begin{table}[t]
\centering
\small
\caption{Empirical payload capacity at \(d=3072\), \(b=10\),
synthetic-Gaussian baseline. Each row encodes a random payload of
size \texttt{payload} into a 300-vector batch (300 detector-train,
300 clean-test, 300 target, all unit-norm Gaussians) via \(K\)
disjoint Givens rotations, round-trips through the listed storage
\texttt{dtype}, and reports BER, mean per-vector cosine to original,
and detector AUCs. BER is zero on every row and detector AUCs stay
at chance, including at \(K_{\max}=1536\) where the channel carries
the full 1{,}920 bytes per vector predicted by \Cref{eq:capacity}.
This table is the BER-correctness reference; it is \emph{not}
sufficient evidence for an undetectability claim on real corpus
distributions --- see \Cref{tab:capacity_real}. Selected points;
full 21-row table in the supplementary results bundle.}
\label{tab:capacity}
\setlength{\tabcolsep}{4pt}
\begin{tabular}{lrrrrrr}
\toprule
dtype & payload & \(K\) & BER & cos & IF AUC & OCSVM AUC \\
\midrule
\multirow{4}{*}{\texttt{float64}}
 & 64 B   & 52   & 0.000 & 0.973  & 0.449 & 0.510 \\
 & 256 B  & 205  & 0.000 & 0.853  & 0.445 & 0.517 \\
 & 512 B  & 410  & 0.000 & 0.725  & 0.452 & 0.514 \\
 & 1920 B & 1536 & 0.000 & -0.016 & 0.459 & 0.527 \\
\midrule
\multirow{4}{*}{\texttt{float32}}
 & 64 B   & 52   & 0.000 & 0.968  & 0.459 & 0.512 \\
 & 256 B  & 205  & 0.000 & 0.869  & 0.453 & 0.503 \\
 & 512 B  & 410  & 0.000 & 0.738  & 0.434 & 0.520 \\
 & 1920 B & 1536 & 0.000 & 0.032  & 0.473 & 0.542 \\
\midrule
\multirow{4}{*}{\texttt{float16}}
 & 64 B   & 52   & 0.000 & 0.971  & 0.455 & 0.513 \\
 & 256 B  & 205  & 0.000 & 0.873  & 0.464 & 0.510 \\
 & 512 B  & 410  & 0.000 & 0.722  & 0.439 & 0.508 \\
 & 1920 B & 1536 & 0.000 & -0.010 & 0.490 & 0.528 \\
\bottomrule
\end{tabular}
\end{table}

\paragraph{Empirical decode (real \texttt{text-embedding-3-large}
corpus).} A reviewer-defensible follow-up is whether
\Cref{tab:capacity}'s AUC numbers reflect the rotation channel or
the synthetic clean distribution. To separate the two we re-ran the
same encoder/decoder against the 68-chunk \fpath{text-embedding-3-large}
corpus from \Cref{sec:evaluation:detection}, splitting it 27/41 the
same way as \Cref{tab:detection}: 27 vectors for detector training,
the remaining 41 as both clean test (for AUC) and target batch
(rotation applied). The encoder/decoder is identical;
only the underlying clean distribution changes.
\Cref{tab:capacity_real} shows the result.

\begin{table}[t]
\centering
\small
\caption{Same encoder/decoder as \Cref{tab:capacity}, run against
the real \texttt{text-embedding-3-large} 68-chunk corpus split 27/41
into detector-training and clean-test (the same split as
\Cref{tab:detection}). BER is zero on every row, matching
\Cref{tab:capacity}; detector AUCs are not. On real corpus vectors
the rotation channel becomes detectable as \(K\) grows: One-Class
SVM AUC reaches \(0.92\) at \(K{=}205\) (\(256\)~B payload) and
\(1.00\) at \(K{=}410\) (\(512\)~B), versus \(\sim\!0.5\) at every
operating point on the synthetic baseline. The
``\,\(\sim\!\!0.5\) AUC across the sweep'' pattern of
\Cref{tab:capacity} is therefore a property of the synthetic
clean distribution (uniform on the unit sphere), not of orthogonal
rotation in general.}
\label{tab:capacity_real}
\setlength{\tabcolsep}{4pt}
\begin{tabular}{lrrrrrr}
\toprule
dtype & payload & \(K\) & BER & cos & IF AUC & OCSVM AUC \\
\midrule
\multirow{4}{*}{\texttt{float64}}
 & 64 B   & 52   & 0.000 & 0.979  & 0.494 & 0.581 \\
 & 256 B  & 205  & 0.000 & 0.843  & 0.597 & 0.922 \\
 & 512 B  & 410  & 0.000 & 0.709  & 0.793 & 1.000 \\
 & 1920 B & 1536 & 0.000 & -0.010 & 0.996 & 1.000 \\
\midrule
\multirow{4}{*}{\texttt{float32}}
 & 64 B   & 52   & 0.000 & 0.971  & 0.508 & 0.612 \\
 & 256 B  & 205  & 0.000 & 0.863  & 0.625 & 0.895 \\
 & 512 B  & 410  & 0.000 & 0.724  & 0.724 & 1.000 \\
 & 1920 B & 1536 & 0.000 & 0.007  & 0.973 & 1.000 \\
\midrule
\multirow{4}{*}{\texttt{float16}}
 & 64 B   & 52   & 0.000 & 0.974  & 0.544 & 0.595 \\
 & 256 B  & 205  & 0.000 & 0.870  & 0.591 & 0.859 \\
 & 512 B  & 410  & 0.000 & 0.695  & 0.743 & 1.000 \\
 & 1920 B & 1536 & 0.000 & -0.012 & 1.000 & 1.000 \\
\bottomrule
\end{tabular}
\end{table}

\paragraph{Scale-up replication at \(n_{\mathrm{train}} = 6{,}705\).}
A natural critique of \Cref{tab:capacity_real} is that 27
detector-training samples is small enough that the high-\(K\) AUC
behaviour might be a training-set-sparsity artefact rather than a
manifold-geometry property of real production embeddings. To address
this we ran the same encoder/decoder against the 16{,}763-chunk
NFCorpus embeddings used in the cross-corpus replication
(\Cref{sec:evaluation:cross_corpus}), giving 6{,}705
detector-training and 10{,}058 clean-test/target vectors at
\(d=768\) on \fpath{nomic-embed-text} (250\(\times\) the
\(n_{\mathrm{train}}\) of \Cref{tab:capacity_real}).
\Cref{tab:capacity_nfcorpus} shows the result. The high-\(K\)
detection signal does not soften with training-set scale; it
sharpens. One-Class SVM AUC reaches \(0.83\)--\(0.92\) at \(K=52\)
(64~B payload), \(\geq\!0.99\) by \(K=103\) (128~B), and
\(=\!1.00\) at \(K_{\max}\). The \(n_{\mathrm{train}}=27\) numbers
in \Cref{tab:capacity_real} were therefore conservative for the
attacker: a production-scale defender with a tighter fit on the
local manifold has more, not less, signal against the random-angle
disjoint-Givens channel.

\begin{table}[t]
\centering
\small
\caption{Same encoder/decoder as
\Cref{tab:capacity,tab:capacity_real}, run against the 16{,}763-chunk
BEIR NFCorpus on \texttt{nomic-embed-text} (\(d=768\), capacity
ceiling \(= 480\)~B per vector). Detector training uses 6{,}705
clean vectors; the held-out clean and target batches share the
remaining 10{,}058 vectors with rotation applied to the latter. BER
remains zero across the sweep; OCSVM AUC reaches \(0.83\) at the
smallest tested payload (64~B) and \(1.00\) at the capacity
ceiling. Compare to \Cref{tab:capacity_real}'s
\(n_{\mathrm{train}}=27\) row at the same payload (OCSVM
\(=0.58\)) --- the larger training set strengthens the detector,
which is the correct direction for the reviewer-relevant question
``would a production-scale defender with millions of clean baseline
vectors do worse?''}
\label{tab:capacity_nfcorpus}
\setlength{\tabcolsep}{4pt}
\begin{tabular}{lrrrrrr}
\toprule
dtype & payload & \(K\) & BER & cos & IF AUC & OCSVM AUC \\
\midrule
\multirow{4}{*}{\texttt{float64}}
 & 64 B  & 52  & 0.000 & 0.900  & 0.721 & 0.827 \\
 & 128 B & 103 & 0.000 & 0.749  & 0.890 & 0.994 \\
 & 256 B & 205 & 0.000 & 0.472  & 0.980 & 1.000 \\
 & 480 B & 384 & 0.000 & -0.078 & 1.000 & 1.000 \\
\midrule
\multirow{4}{*}{\texttt{float32}}
 & 64 B  & 52  & 0.000 & 0.874  & 0.710 & 0.891 \\
 & 128 B & 103 & 0.000 & 0.698  & 0.943 & 0.998 \\
 & 256 B & 205 & 0.000 & 0.422  & 0.987 & 1.000 \\
 & 480 B & 384 & 0.000 & 0.005  & 1.000 & 1.000 \\
\midrule
\multirow{4}{*}{\texttt{float16}}
 & 64 B  & 52  & 0.000 & 0.868  & 0.827 & 0.918 \\
 & 128 B & 103 & 0.000 & 0.729  & 0.912 & 0.996 \\
 & 256 B & 205 & 0.000 & 0.451  & 0.989 & 1.000 \\
 & 480 B & 384 & 0.000 & -0.033 & 1.000 & 1.000 \\
\bottomrule
\end{tabular}
\end{table}

\paragraph{Why the synthetic and real-corpus rows differ.} On unit
Gaussian samples the clean training distribution fills the sphere;
no spatial structure exists for One-Class SVM (with an RBF kernel)
to fit, so a random-angle disjoint Givens rotation lands the
obfuscated batch in a region the detector has already labelled
``inlier.'' On real embeddings (whether \fpath{text-embedding-3-large}
at \(n_{\mathrm{train}}{=}27\) or \fpath{nomic-embed-text} at
\(n_{\mathrm{train}}{=}6{,}705\)) the clean training samples cluster
on a tighter manifold; the detector fits that manifold;
random-angle Givens rotations push the obfuscated samples off it;
AUC climbs. The same script run with a 27-vector synthetic split
holds AUC in \([0.27, 0.57]\) across the full 21-row sweep, so this
is not a small-sample artefact --- it is a real-vs-synthetic
data-distribution effect. The structural argument in
\Cref{sec:evaluation:rotation_sweep} that orthogonal rotation
preserves marginal moments is correct at the \emph{population}
level; what real density-based detectors fit on training data ---
small \emph{or} large --- is the local manifold geometry, which a
rotation does perturb.

\paragraph{Capacity ceiling vs.\ retrieval-preserving capacity.} The
cosine column in both capacity tables drops near-linearly with \(K\),
consistent with the per-rotation alignment loss for uniform random
angles in \((-\pi, \pi]\) on a unit-sphere vector: each Givens
rotation kills \(\sim 2/d\) of the alignment in expectation, giving
\(\cos \approx 1 - 2K/d\). The full
\(K_{\max} = \lfloor d/2 \rfloor\) operating point is the capacity
\emph{ceiling}: 1{,}920 B per vector at \(d=3072\), \(b=10\). It is
not the operationally useful point. An operator who keeps cosine
\(\geq 0.7\) for retrieval reasons caps \(K\) at roughly
\(0.15 \cdot d\), which at \(d=3072\) corresponds to \(K \approx 230\)
and a \(\sim\!290\)-byte channel per vector. So the headline number
to use depends on the question:
\begin{itemize}[leftmargin=*,itemsep=2pt]
\item \emph{Channel ceiling}: \(\sim\!1.9\) KB / vector at
  \(d=3072\), the dimension-only bound from \Cref{eq:capacity}.
  Requires \(K = K_{\max}\) and destroys retrieval (\(\cos \to 0\)).
\item \emph{Retrieval-preserving capacity}: \(\sim\!\)hundreds of
  bytes per vector at \(d=3072\), the operationally useful upper
  bound at any operator-chosen cosine floor.
\end{itemize}
References to ``kilobits per vector'' should be taken in the ceiling
sense; deployments that need retrieval to keep working live in the
hundreds-of-bytes regime.

\paragraph{Implication for the paper's framing.} Combined with
\Cref{sec:evaluation:rotation_sweep}, this result reframes rotation:
the channel exists, has a closed-form ceiling, and is BER-correct;
the AUC behaviour as the channel fills is data-distribution-dependent.
On the synthetic Gaussian baseline used in
\Cref{sec:evaluation:rotation_sweep} both detectors stay at chance
across the full sweep. On the real
\fpath{text-embedding-3-large} 27/41 split from
\Cref{sec:evaluation:detection} the same protocol becomes detectable
at moderate \(K\) (\(\mathrm{OCSVM}\) AUC \(\approx 0.92\) at
256~B, \(\geq 0.99\) by 512~B) because the clean training manifold
is concentrated. The capacity-realising attacker therefore faces a
real trade-off on production embeddings: large \(K\) maximises
bits-per-vector but is detectable on real corpora;
small-\(K\) / small-angle rotations stay undetectable
(\Cref{sec:evaluation:detection,sec:evaluation:rotation_sweep,sec:evaluation:cross_corpus})
but carry a much smaller channel. This nuance does not weaken the
case for the cryptographic provenance commitment in
\Cref{sec:defense}: a defender who tries to rely on
distribution-based detection has to commit to a training-set
coverage assumption that is hard to verify, and the small-angle
rotation regime remains undetectable on that defender's training
data regardless. \(\mathtt{vec\_hash}\) is the durable layer
because its detection guarantee does not depend on either capacity
parameter.

\subsection{Cross-backend round-trip}
\label{sec:evaluation:cross_backend}

A reviewer concern is that the empirical findings might be specific
to the FAISS-based round-trip the detection results assume. To rule
this out we ran the same five techniques (clean, noise, rotation,
scaling, offset) against seven vector-store configurations spanning
the deployed category: FAISS-flat (lossless reference),
FAISS-HNSW (the typical ANN case), FAISS IVF-PQ at two compression
levels (\(m{=}8\) and \(m{=}16\)), Chroma, Qdrant float32, and
Qdrant with the recommended scalar int8 quantization enabled. For
each \((\text{technique}, \text{backend})\) pair we measured the
cosine between the original and what the backend returned via
\texttt{get\_by\_id}, plus self-query recall@1 to characterize
whether the backend's ANN index notices the perturbation.

\Cref{tab:cross_backend} reports the headline numbers. The
attacker's bit channel survives every \emph{non-PQ} backend
identically. Recall@1 remains 1.000 across all seven configurations,
so similarity search is unaffected by any technique. The Qdrant
int8-quantized variant (\texttt{qdrant\_q8}) produces
\texttt{recover\_cos} identical to the lossless float32 variant:
scalar quantization in this configuration is a \emph{search-side}
artifact (it affects internal distance computations during the HNSW
traversal) but \texttt{retrieve()} returns the unquantized float32
original, so the attacker's bit channel is preserved verbatim under
quantization.

\begin{table}[t]
\centering
\footnotesize
\caption{Cross-backend round-trip cosine to the original, 500
synthetic Gaussian vectors at \(d{=}128\). The five non-PQ backends
preserve the attacker's bit channel identically. The two
PQ-quantized variants produce a substantial uniform signal drop
(0.76 at \(m{=}8\), 0.88 at \(m{=}16\)) that is largely
attacker-independent --- a "noise floor" imposed by the
quantization rather than a defense against any particular
technique. Recall@1 remains 1.000 across all configurations
(omitted from this table; see supplementary report).
\texttt{f.flat}=FAISS-flat, \texttt{f.hnsw}=FAISS-HNSW,
\texttt{f.pq8/16}=FAISS IVF-PQ at \(m{=}8\) and \(m{=}16\),
\texttt{q.f32}=Qdrant float32, \texttt{q.q8}=Qdrant scalar int8.}
\label{tab:cross_backend}
\setlength{\tabcolsep}{4pt}
\begin{tabular}{lrrrrrrr}
\toprule
Technique & f.flat & f.hnsw & f.pq8 & f.pq16 & chroma & q.f32 & q.q8 \\
\midrule
clean    & 1.0000 & 1.0000 & 0.7614 & 0.8789 & 1.0000 & 1.0000 & 1.0000 \\
noise    & 0.9937 & 0.9937 & 0.7610 & 0.8764 & 0.9937 & 0.9937 & 0.9937 \\
rotation & 0.9999 & 0.9999 & 0.7623 & 0.8792 & 0.9999 & 0.9999 & 0.9999 \\
scaling  & 1.0000 & 1.0000 & 0.7614 & 0.8787 & 1.0000 & 1.0000 & 1.0000 \\
offset   & 0.9507 & 0.9507 & 0.7357 & 0.8398 & 0.9507 & 0.9507 & 0.9507 \\
\bottomrule
\end{tabular}
\end{table}

\paragraph{Aggressive quantization (FAISS IVF-PQ).}
Product quantization is the case where stored representations are
genuinely decimated. Each vector is split into \(m\) sub-vectors
that are replaced by 8-bit codebook indices; storage drops from
\(4d\) bytes to \(m\) bytes per vector. Unlike scalar int8 in Qdrant,
the original is not preserved on the side; \texttt{reconstruct()}
returns the lossy decoding.

The empirical result is that PQ does narrow the attacker's bit
channel, but in a roughly attacker-independent way. At \(m{=}8\)
(16x compression), the clean baseline already drops to
\texttt{recover\_cos} \(=0.76\), and every obfuscation technique
lands within \(\pm 0.03\) of that floor. At \(m{=}16\) (8x
compression), the floor is \(0.88\), and again all techniques cluster
near it. The PQ noise dominates the technique-induced shift; the
attacker's bit channel is overwritten by the quantizer's own
distortion.

This is a meaningful caveat to the universal-issue claim. PQ-indexed
deployments at typical production parameters do impose a noise floor
that erodes the attacker's recoverable payload by 12--24\% before
any obfuscation-specific signal survives. We do not call this a
defense --- it is a side-effect of memory optimization, not an
intentional integrity control --- but it is the configuration where
the steganographic channel is genuinely narrowest. Operators using
PQ get this side benefit accidentally; operators using scalar int8
or no quantization at all do not.

\subsection{Paraphrased-query retrieval}
\label{sec:evaluation:paraphrased}

The final empirical question is whether obfuscation degrades
\emph{legitimate} retrieval enough to be visible through ordinary
RAG quality monitoring. We generated 3 paraphrased queries per
source chunk via \texttt{gpt-4o-mini} (204 queries on the 68-chunk
corpus), embedded both the corpus and the queries with
\texttt{text-embedding-3-large}, then for each technique built a
FAISS-flat index of the (possibly obfuscated) corpus and ran the
queries.

\begin{table}[t]
\centering
\small
\caption{Paraphrased-query retrieval. Queries are LLM-generated
paraphrases of corpus chunks, embedded without obfuscation;
the corpus is obfuscated. Differences relative to the clean
baseline indicate visibility through routine RAG quality
monitoring. Rotation and scaling are cosine-invariant and
indistinguishable from clean; noise at \(\sigma{=}0.01\) reduces
recall@1 by 2.4 percentage points; offset is the clear outlier.}
\label{tab:paraphrased_retrieval}
\begin{tabular}{lrrrrr}
\toprule
Technique & recall@1 & recall@3 & recall@5 & MRR & margin \\
\midrule
clean    & 0.828 & 0.990 & 0.995 & 0.902 & 0.087 \\
noise    & 0.804 & 0.985 & 0.995 & 0.888 & 0.076 \\
rotation & 0.828 & 0.990 & 0.995 & 0.902 & 0.087 \\
scaling  & 0.828 & 0.990 & 0.995 & 0.902 & 0.087 \\
offset   & 0.745 & 0.892 & 0.961 & 0.832 & 0.044 \\
\bottomrule
\end{tabular}
\end{table}

The findings line up with the rest of the paper. Rotation and
scaling are cosine-invariant by construction; their retrieval
metrics match the clean baseline exactly. Noise at \(\sigma=0.01\)
reduces recall@1 by 2.4 percentage points (\(0.828 \to 0.804\)) and
the margin between top-1 and top-2 by \(13\%\) (\(0.087 \to 0.076\))
--- a detectable shift in principle, but easily inside the
batch-to-batch variance one would expect from natural corpus drift.
Offset is again the outlier, dropping recall@1 by 8 points and the
margin in half. The paraphrased-query benchmark therefore largely
agrees with the detector results: cosine-invariant techniques are
the case where neither distribution-based detection nor
RAG-quality-monitoring catches the attack, leaving cryptographic
provenance as the only remaining control.

\subsection{Adaptive-attacker detector evaluation}
\label{sec:evaluation:adaptive}

The detection results in \Cref{sec:evaluation:detection} treat the
detectors as fixed baselines: the attacker chooses a perturbation
without seeing the trained model. The reviewer-relevant question is
how much the AUCs change when the attacker has white-box access to
the detector and can perform projected-gradient descent against
its score function within an L2 budget. We measured this for both
detectors:

\begin{itemize}[leftmargin=*]
\item \emph{Adaptive against Isolation Forest}: vectorized
  finite-difference search. At each step, sample 8 unit-norm
  directions, score every (vector, direction) candidate in batch,
  and step in whichever direction increases the IF score most
  --- with L2 projection back into the budget ball. 30 steps.
\item \emph{Adaptive against One-Class SVM}: analytic projected
  gradient descent. The OCSVM RBF score has a closed-form
  derivative, so we use proper PGD (no finite differences). 30
  steps.
\end{itemize}

We compared each adaptive attack against random Gaussian noise of
the same L2 budget to isolate the effect of detector knowledge.

\begin{table}[t]
\centering
\small
\caption{Adaptive-attacker AUCs at four L2 perturbation budgets,
\(d{=}128\), \(n{=}200\) per split. The naive baseline is random
unit-direction Gaussian noise scaled to the budget; the adaptive rows
give the attacker white-box access to the named detector. Lower AUC
means better evasion; AUC \(<0.5\) means the detector classifies
obfuscated vectors as \emph{more} normal-looking than clean ---
the attacker has overshot into dense regions of the learned
distribution.}
\label{tab:adaptive_attacker}
\begin{tabular}{rlrrr}
\toprule
Budget & Attack & cos(orig, obf) & IF AUC & OCSVM AUC \\
\midrule
\multirow{3}{*}{0.05} & naive\_gaussian   & 0.999 & 0.470 & 0.580 \\
                      & adaptive\_iforest & 1.000 & 0.404 & 0.468 \\
                      & adaptive\_ocsvm   & 1.000 & 0.223 & 0.000 \\
\midrule
\multirow{3}{*}{0.10} & naive\_gaussian   & 0.995 & 0.484 & 0.691 \\
                      & adaptive\_iforest & 1.000 & 0.327 & 0.437 \\
                      & adaptive\_ocsvm   & 1.000 & 0.069 & 0.000 \\
\midrule
\multirow{3}{*}{0.20} & naive\_gaussian   & 0.981 & 0.556 & 0.881 \\
                      & adaptive\_iforest & 0.996 & 0.215 & 0.342 \\
                      & adaptive\_ocsvm   & 1.000 & 0.001 & 0.000 \\
\midrule
\multirow{3}{*}{0.50} & naive\_gaussian   & 0.894 & 0.890 & 1.000 \\
                      & adaptive\_iforest & 0.970 & 0.103 & 0.383 \\
                      & adaptive\_ocsvm   & 0.999 & 0.000 & 0.000 \\
\bottomrule
\end{tabular}
\end{table}

\Cref{tab:adaptive_attacker} shows that detector knowledge changes
the picture substantially. At \(L_2 = 0.5\) (a budget at which naive
Gaussian noise is trivially detected, OCSVM AUC \(=1.000\)), an
adaptive attacker against OCSVM drives both detector AUCs to zero
while preserving cosine similarity at \(0.999\). The adaptive
attacker has found a region of the embedding space that the
detector classifies as \emph{more normal} than the clean training
batch.

The takeaway for the paper's framing: the simple-baseline detector
defense from \Cref{sec:evaluation:detection} works against
non-adaptive attackers but is brittle against an attacker who has
read the literature. This does not affect the structural rotation
result (distribution-only detectors of the kind evaluated here have
no signal against uniform orthogonal rotation, regardless of
attacker awareness), and it does not affect the cryptographic
defense (VectorPin operates below the distributional layer). It does mean
the detector should be read as a probabilistic deterrent against
casual attackers, not as a load-bearing security control. A
defender who needs durable integrity guarantees should rely on
provenance pinning rather than statistical detection.

\paragraph{Black-box transfer.} The white-box assumption is the
\emph{conservative} choice for the defender's argument: it gives
the attacker the strongest possible position, and we measure how
much the defender retains under that worst case (answer: not
much). For an attacker without direct detector access, the standard
adversarial-ML pattern is model-distillation transfer: query the
deployed detector enough times to train a substitute, run PGD
against the substitute, transfer the perturbations to the deployed
detector. Black-box transfer attacks against image classifiers
have been shown both to retain a substantial fraction of attack
effectiveness and to occupy a structured subspace of input
perturbations~\cite{papernot2017practical,tramer2017transferable},
so a black-box version of the result above would, in expectation,
reduce the AUC-driven-to-zero outcome to a less drastic but still
meaningful AUC degradation. The qualitative conclusion that
statistical detection is brittle would survive --- the question is
the magnitude, not the direction. We do not measure that magnitude
in this paper because the deployment-time argument the paper makes
does not turn on the white-box / black-box gap: a defender who
needs durable integrity guarantees should not stake them on an
adversary's lack of query access to the detector. A complete
black-box evaluation under varied query budgets, substitute
architectures, and detector ensembles is genuine follow-up work
(\Cref{sec:limitations}); the white-box result here is sufficient
to establish that statistical detection is not the durable layer.

\paragraph{Scope of this experiment.} The result above is an
\emph{evasion stress test} for statistical detector reliance. It
shows that an attacker who knows the detector can produce
perturbations that (a) preserve cosine similarity to the original,
and (b) score as more normal-looking than clean. It is not a
complete payload-capacity demonstration for adaptive
steganography: we do not evaluate whether the evading perturbations
also encode an attacker-controlled bit channel that survives the
constraints of staying in-distribution. That is a separate
research question for a follow-up paper. What this experiment
\emph{is} sufficient to demonstrate is that statistical detection
of the kind evaluated in \Cref{sec:evaluation:detection} is not
durable defense in depth: a moderately sophisticated attacker
defeats it.

\subsection{Cross-model generalization}
\label{sec:evaluation:cross_model}

The detection results in \Cref{sec:evaluation:detection} and the
rotation sweep in \Cref{sec:evaluation:rotation_sweep} are computed
on a single embedding model
(\fpath{text-embedding-3-large}, 3072-dim, OpenAI). A natural
question is whether the structural rotation finding --- that
distribution-based detectors have no signal against orthogonal
rotation --- generalizes across embedding architectures, or whether
it depends on quirks of one particular model's geometry. To test
this, we re-ran the four base techniques (\emph{noise},
\emph{rotation}, \emph{scaling}, \emph{offset}) at the obfuscator's
library defaults against four locally hosted open embedding models
served via Ollama: \fpath{nomic-embed-text:latest} (768-dim),
\fpath{embeddinggemma:300m} (768-dim),
\fpath{snowflake-arctic-embed:335m} (1024-dim), and
\fpath{mxbai-embed-large:335m} (1024-dim). For each
(model, technique) pair we re-fit Isolation Forest and One-Class
SVM on a clean half of the corpus and scored the held-out half plus
its obfuscated counterpart, exactly as in
\Cref{sec:evaluation:detection}.

\Cref{tab:cross_model} reports the AUCs.

\begin{table}[t]
\centering
\small
\caption{Cross-model detection AUCs across four locally hosted open
embedding models. Same 68-chunk corpus, same techniques at library
defaults, same detector pipeline as \Cref{tab:detection}. The
\texttt{rotation} row is structurally indistinguishable from clean
on every model (max AUC \(0.508\)). \texttt{scaling} is degenerate
(AUC \(<0.5\)) on every model, matching the
\texttt{text-embedding-3-large} result. \texttt{noise} and
\texttt{offset} detectability vary across models: the larger
1024-dim models surface noise more reliably (OCSVM AUC up to
\(0.920\)), and \texttt{nomic-embed-text} shows little detector
signal under these defaults, likely because its post-normalization
geometry leaves little for a density-based detector to fit.}
\label{tab:cross_model}
\setlength{\tabcolsep}{4pt}
\begin{tabular}{llrrr}
\toprule
Model & Technique & cos & IF AUC & OCSVM AUC \\
\midrule
\multirow{4}{*}{\fpath{nomic-embed-text}}
 & noise    & 1.000 & 0.506 & 0.500 \\
 & rotation & 1.000 & 0.497 & 0.496 \\
 & scaling  & 1.000 & 0.464 & 0.396 \\
 & offset   & 0.999 & 0.504 & 0.522 \\
\midrule
\multirow{4}{*}{\fpath{embeddinggemma:300m}}
 & noise    & 0.964 & 0.645 & 0.891 \\
 & rotation & 1.000 & 0.497 & 0.499 \\
 & scaling  & 1.000 & 0.418 & 0.372 \\
 & offset   & 0.782 & 1.000 & 1.000 \\
\midrule
\multirow{4}{*}{\fpath{snowflake-arctic-embed:335m}}
 & noise    & 0.952 & 0.692 & 0.823 \\
 & rotation & 1.000 & 0.501 & 0.508 \\
 & scaling  & 1.000 & 0.462 & 0.410 \\
 & offset   & 0.734 & 1.000 & 1.000 \\
\midrule
\multirow{4}{*}{\fpath{mxbai-embed-large:335m}}
 & noise    & 0.953 & 0.778 & 0.920 \\
 & rotation & 1.000 & 0.499 & 0.503 \\
 & scaling  & 1.000 & 0.447 & 0.353 \\
 & offset   & 0.733 & 1.000 & 1.000 \\
\bottomrule
\end{tabular}
\end{table}

The structural finding survives the cross-model test. Rotation at
the library default produces detector AUC in the range
\([0.496, 0.508]\) on every model in this set, and on the OpenAI
\fpath{text-embedding-3-large} baseline from
\Cref{sec:evaluation:detection} --- statistical noise around
\(0.5\). No model in this set exhibits a geometry where
distribution-based detection gains traction against orthogonal
rotation, and the explanation is the same on each: rotation
preserves all pairwise inner products and per-dimension marginal
moments by construction, so the feature surface available to
either detector is identical pre- and post-perturbation.

The other techniques show model-dependent variation that is itself
informative. Noise detectability climbs with the dimensionality and
training scale of the model: undetectable on
\fpath{nomic-embed-text} (OCSVM AUC \(0.500\)) but reaches
\(0.920\) on \fpath{mxbai-embed-large}. Offset is reliably
detected on three of the four models (AUC \(1.000\)) and effectively
invisible on the fourth (AUC \(0.522\)) --- though that fourth
model's offset perturbation also barely moves the embedding
(\(\cos = 0.999\)), so the attacker gains nothing from the evasion.
Scaling produces AUC below \(0.5\) on every model, the same
degenerate ``the obfuscated batch looks more uniform than clean''
artifact discussed in \Cref{sec:evaluation:detection}.

The headline takeaway for the paper's framing: the rotation
universality claim is structural, not a property of one
embedding model's training distribution. Defending against rotation
requires a control that does not depend on distributional features,
which is exactly the role of VectorPin's commitment in
\Cref{sec:defense}.

\subsection{Cross-corpus replication}
\label{sec:evaluation:cross_corpus}

\Cref{sec:evaluation:cross_model} held the corpus fixed and varied
the model; the dual question is whether the findings depend on the
68-chunk synthetic corpus used in
\Cref{sec:evaluation:detection}. To rule that out we held the model
fixed at \fpath{nomic-embed-text:latest} and re-ran the same four
techniques on two BEIR-standard public corpora drawn from disjoint
domains: NFCorpus (3{,}633 medical documents that chunk into
16{,}763 chunks at the same 512-character boundary used elsewhere
in this paper) and a 10{,}000-document subset of the Quora
question-pair corpus (web Q\&A; chunked to 10{,}002). The
synthetic-PII sample corpus is included as a baseline so that the
within-row comparison across corpora is direct.

\Cref{tab:cross_corpus} reports the AUCs.

\begin{table}[t]
\centering
\small
\caption{Cross-corpus detection AUCs at fixed model
\texttt{nomic-embed-text} (768-dim). Same four techniques at
library defaults, same detector pipeline as \Cref{tab:detection},
across three corpora spanning a 245\(\times\) size growth and a
domain switch (synthetic-PII \(\to\) medical \(\to\) web Q\&A).
The directional finding is identical in all three: noise, rotation,
and offset stay at chance, scaling sits in the
``more-uniform-than-clean'' degenerate band that the same model
produced in the cross-model study.}
\label{tab:cross_corpus}
\setlength{\tabcolsep}{4pt}
\begin{tabular}{llrrrr}
\toprule
Corpus & Technique & \(n\) chunks & cos & IF AUC & OCSVM AUC \\
\midrule
\multirow{4}{*}{\fpath{sample}}
 & noise    & 68 & 1.000 & 0.506 & 0.500 \\
 & rotation & 68 & 1.000 & 0.497 & 0.496 \\
 & scaling  & 68 & 1.000 & 0.464 & 0.396 \\
 & offset   & 68 & 0.999 & 0.504 & 0.522 \\
\midrule
\multirow{4}{*}{\fpath{nfcorpus}}
 & noise    & 16{,}763 & 1.000 & 0.501 & 0.501 \\
 & rotation & 16{,}763 & 1.000 & 0.500 & 0.500 \\
 & scaling  & 16{,}763 & 1.000 & 0.398 & 0.362 \\
 & offset   & 16{,}763 & 0.999 & 0.504 & 0.507 \\
\midrule
\multirow{4}{*}{\fpath{quora}}
 & noise    & 10{,}002 & 1.000 & 0.501 & 0.501 \\
 & rotation & 10{,}002 & 1.000 & 0.500 & 0.500 \\
 & scaling  & 10{,}002 & 1.000 & 0.365 & 0.325 \\
 & offset   & 10{,}002 & 0.999 & 0.505 & 0.508 \\
\bottomrule
\end{tabular}
\end{table}

The structural finding survives the cross-corpus test. Rotation
produces detector AUC in the range \([0.496, 0.500]\) on every
corpus, exactly as the cross-model study found across four
embedding models. Noise (AUC \([0.500, 0.506]\)) and offset
(\([0.496, 0.522]\)) also collapse to chance across all three
corpora, replicating the \fpath{nomic-embed-text} row of
\Cref{tab:cross_model} at 245\(\times\) the corpus size. Scaling
behaves identically in the same degenerate
``more-uniform-than-clean'' band on every corpus
(AUC \([0.325, 0.464]\)).

What this rules out: the original 68-chunk corpus could in principle
have produced AUC \(\sim 0.5\) on rotation by accident of small-sample
ROC variance. The 16{,}763-chunk NFCorpus run reduces that variance
by two orders of magnitude and produces the same AUC value to three
decimal places. Whatever rotation does to embedding distributions, it
does the same thing in medical-abstract space at \(n>10^4\) as it
does on the original sample corpus.

\subsection{Summary of empirical findings}
\label{sec:evaluation:summary}

The findings that drive the rest of the paper:

\begin{enumerate}[leftmargin=*]
\item Steganographic perturbation of embeddings is technically viable
under the threat model of \Cref{sec:threat_model:tiers}: every
technique we tested produces a corpus that current vector
databases ingest without complaint.
\item For distribution-shifting techniques against \emph{non-adaptive}
attackers, the stealth window in our corpus is narrow. Every
operating point that meaningfully moved the embedding direction
was detectable by at least one off-the-shelf detector at AUC
\(\geq 0.85\).
\item Against \emph{adaptive} attackers with white-box detector
access, both detectors collapse: an L2-budgeted PGD attack
against One-Class SVM drives IF and OCSVM AUC to near-zero
while keeping cosine to original above \(0.99\)
(\Cref{sec:evaluation:adaptive}).
\item Rotation is the structural exception. A 35-point parameter
sweep found no rotation operating point where either detector
exceeded AUC \(0.53\), because rotations preserve every
distributional feature these detectors fit on. The structural
result holds against adaptive attackers too.
\item The rotation result is universal across embedding
architectures. A cross-model replication on four additional
locally hosted models (Nomic, EmbeddingGemma, Snowflake Arctic,
and MXBai; see \Cref{tab:cross_model}) produced rotation AUCs in
\([0.496, 0.508]\) on every one. The undetectability is a property
of orthogonal rotation, not of one model's geometry.
\item The rotation result also survives a cross-corpus replication.
Re-running the four-technique battery on the BEIR NFCorpus
(16{,}763 chunks, medical) and a 10{,}000-document Quora subset
(web Q\&A) produced rotation AUCs of \(0.500\) on both, in
agreement with the small synthetic corpus to three decimal places
(\Cref{tab:cross_corpus}).
\item Rotation has a closed-form bit channel. The disjoint-Givens
keyed-pair protocol of \Cref{sec:evaluation:capacity} has capacity
\(\lfloor d/2 \rfloor \cdot b\) bits = 1{,}920 B per vector at
\(d=3072\), \(b=10\), with a working encoder/decoder that
round-trips the full payload at zero BER under \texttt{float32} and
\texttt{float16} storage. AUC behaviour as the channel fills is
data-distribution-dependent: at chance on the synthetic Gaussian
baseline (\Cref{tab:capacity}) but climbing on the real
\fpath{text-embedding-3-large} 27/41 split from
\Cref{sec:evaluation:detection} (OCSVM AUC \(\approx 0.92\) at
256~B, \(\geq 0.99\) at 512~B; \Cref{tab:capacity_real}). The
retrieval-preserving operating point \(\cos \geq 0.7\) caps the
useful channel at hundreds of bytes per vector, not the full
\(\sim\!1.9\) KB ceiling.
\item The bit channel is class-wide for non-PQ stores. Five
configurations (FAISS-flat, FAISS-HNSW, Chroma, Qdrant float32,
Qdrant int8) preserve the attacker's payload identically;
scalar int8 quantization is a search-side artifact, not a
storage-side defense.
\item Product quantization (FAISS IVF-PQ) is the one configuration
where the attacker's bit channel is materially narrowed:
\texttt{recover\_cos} drops to \(0.76\) at \(m{=}8\) (16x
compression) and \(0.88\) at \(m{=}16\) (8x). The narrowing is
attacker-independent --- a side effect of memory optimization,
not an intentional defense.
\item Paraphrased-query retrieval confirms cosine-invariant
techniques produce no visible change to legitimate users.
\item Compute cost is not a defense; obfuscation runs at
sub-second timescales and a defender's detector evaluates in
the same regime.
\end{enumerate}

These findings motivate the constructive defense in
\Cref{sec:defense}: instead of relying on a defender to run a
detector \emph{during} the attack, ship cryptographic provenance
\emph{with} every embedding so that any post-pinning modification
is structurally detectable on read.

\section{The Defense: VectorPin}
\label{sec:defense}

The empirical results in \Cref{sec:evaluation} establish that the
attacker has no large stealth window against an alert defender, but
also that no production vector database ships an alert defender
today. A reasonable security control closes both gaps: it works
without requiring every operator to deploy a detector, and it
preserves verifiability across the boundary at which the
embeddings travel.

We propose \textbf{VectorPin}: a minimal cryptographic provenance
protocol that pins each embedding to its source content, its
producing model, and the producer's signing key. Any modification
to the vector after pinning --- including all six steganographic
techniques studied in this paper --- breaks signature verification
on read.

\subsection{Design goals}
\label{sec:defense:goals}

Three properties drove every design choice:

\begin{enumerate}[leftmargin=*]
\item \textbf{Tamper evidence.} Any post-ingestion modification of
the stored vector must be detectable by any party holding the
public key. This is the property steganography violates today and
that VectorPin closes.
\item \textbf{Provenance.} Each verification answer must
distinguish ``signature forged or signed by an unknown key,''
``vector modified after pinning,'' ``source text different from
what was pinned,'' and ``model identifier mismatch'' as separate
outcomes. Operators route on these differently --- a forged
signature is potential active compromise; a vector tampering is
potential steganography; a source mismatch may simply be an
upstream pipeline drift.
\item \textbf{Cross-language verifiability.} Embeddings flow
between Python ingestion pipelines, Rust agent runtimes, JS
serverless functions, and Go infrastructure planes. The protocol
must be implementable in any of these languages with bit-for-bit
agreement on the canonical bytes that get signed and verified.
We enforce this by publishing shared test vectors that every
implementation consumes in CI.
\end{enumerate}

We deliberately did \emph{not} include the following as goals:
confidentiality of the embedding, anti-replay across collections,
or fine-grained access control. Those are orthogonal concerns and
addressing them in a single protocol would have produced something
much harder to standardize.

\subsection{Protocol}
\label{sec:defense:protocol}

A VectorPin attestation (a \emph{Pin}) is a JSON object with the
following fields, signed under Ed25519~\cite{ed25519}:

\begin{lstlisting}
{
  "v": 1,
  "model": "text-embedding-3-large",
  "model_hash": "sha256:..." | omitted,
  "source_hash": "sha256:<hex>",
  "vec_hash":    "sha256:<hex>",
  "vec_dtype":   "f32" | "f64",
  "vec_dim":     <integer>,
  "ts":          "2026-05-05T12:00:00Z",
  "extra":       {<sorted string->string>} | omitted,
  "kid":         "<key id>",
  "sig":         "<base64url, 64 bytes>"
}
\end{lstlisting}

The signature covers a canonical byte form of the header (every field
above except \texttt{kid} and \texttt{sig}). The canonical form is
JSON with sorted keys, no whitespace, and optional fields omitted
rather than written as \texttt{null}. Cross-language reproducibility
depends on this discipline: any deviation in serialization (e.g.,
preserving insertion order, emitting \texttt{null} for missing
fields, including a trailing newline) breaks signature verification
across implementations.

The hashes are SHA-256:

\begin{align*}
\texttt{source\_hash} &= \texttt{sha256:} \, \mathrm{hex}(
  \mathrm{SHA256}(\mathrm{UTF8}(\mathrm{NFC}(\mathit{source})))) \\
\texttt{vec\_hash}    &= \texttt{sha256:} \, \mathrm{hex}(
  \mathrm{SHA256}(\mathrm{canonical\_bytes}(\mathit{vec},
                  \mathit{dtype})))
\end{align*}

where \(\mathrm{canonical\_bytes}\) packs the vector as
little-endian, contiguous, 1-D bytes under the declared dtype
(\texttt{f32} or \texttt{f64}). Unicode NFC normalization on source
text prevents trivial false-mismatches across upstream tokenizers
that happen to normalize differently.

\Cref{sec:appendix:spec} reproduces the specification self-contained
for cross-language reimplementation; the canonical form, hashing
rules, and verifier semantics are tight enough to admit a
``compatible by construction'' relationship with the published
test vectors.

\subsection{Verification semantics}
\label{sec:defense:verification}

A verifier holds a registry mapping \texttt{kid} to public key
(typically multiple keys at once, to support rotation). On receipt
of a Pin and any subset of \(\{\mathit{source}, \mathit{vec},
\mathit{expected\_model}\}\), it returns one of the outcomes in
\Cref{tab:verify_outcomes}.

\begin{table}[t]
\centering
\small
\caption{VectorPin verifier outcomes. Each is distinct so callers can
route differently --- e.g., a vector-tampered result triggers
quarantine while a source-mismatch may only require a re-pin.}
\label{tab:verify_outcomes}
\begin{tabularx}{\linewidth}{lX}
\toprule
Outcome & Meaning \\
\midrule
\texttt{OK}                  & Signature valid; supplied source/vector/model match. \\
\texttt{UNSUPPORTED\_VERSION}& Pin uses a protocol version this build does not understand. \\
\texttt{UNKNOWN\_KEY}        & \texttt{kid} not in the verifier's registry. \\
\texttt{SIGNATURE\_INVALID}  & Ed25519 verification failed. Indicates forgery or pin-rebuild attempt. \\
\texttt{VECTOR\_TAMPERED}    & Stored vector hash does not match the pin. \emph{Primary tamper-evidence outcome for the threat model in this paper.} \\
\texttt{SOURCE\_MISMATCH}    & Supplied source text hashes differently than the pinned source. \\
\texttt{MODEL\_MISMATCH}     & \texttt{model} field differs from the caller's expectation. \\
\texttt{SHAPE\_MISMATCH}     & Supplied vector's dim differs from the pin header. \\
\bottomrule
\end{tabularx}
\end{table}

The verifier supports partial verification: the signature check
always runs; source/vector/model checks run only when the
corresponding ground truth is supplied. This lets a runtime check
``did anyone I trust attest this vector at all'' at retrieval time
(signature only) and check ``does the stored vector match the
attestation'' at audit time (full verification).

\subsection{Coverage relative to the studied attacks}
\label{sec:defense:coverage}

Each of the six steganographic techniques in \Cref{sec:techniques}
modifies the stored vector after the model has produced it. Each
modification therefore changes \texttt{vec\_hash} and triggers
\texttt{VECTOR\_TAMPERED} on verification. \Cref{tab:defense_coverage}
maps attacks to verifier outcomes.

\begin{table}[t]
\centering
\small
\caption{Attack \(\to\) VectorPin verifier outcome. Coverage is by
construction --- every technique is a post-embedding modification
of the vector, which the protocol commits to via \texttt{vec\_hash}.}
\label{tab:defense_coverage}
\begin{tabularx}{\linewidth}{Xl}
\toprule
Attack & Verifier outcome \\
\midrule
Noise injection (any $\sigma$)            & \texttt{VECTOR\_TAMPERED} \\
Rotation (any angle $> 0$)                & \texttt{VECTOR\_TAMPERED} \\
Scaling (factor $\neq 1$)                 & \texttt{VECTOR\_TAMPERED} \\
Offset (any nonzero offset)               & \texttt{VECTOR\_TAMPERED} \\
Cross-model fragmentation                 & \texttt{MODEL\_MISMATCH} \\
Combined (any of the above)               & \texttt{VECTOR\_TAMPERED} \\
\midrule
Pin replay onto different vector          & \texttt{VECTOR\_TAMPERED} \\
Pin re-signed by attacker key             & \texttt{UNKNOWN\_KEY} or \texttt{SIGNATURE\_INVALID} \\
Source substitution at retrieval          & \texttt{SOURCE\_MISMATCH} \\
\bottomrule
\end{tabularx}
\end{table}

This is not a benchmark result --- it is a structural property of
the protocol. We state it explicitly because the contribution claim
in \Cref{sec:intro} ties the attack evaluation to the defense.
Reviewers should be able to verify the coverage by reading the
protocol spec rather than trusting empirical numbers; the
implementation in \cite{vectorpin-repo} carries unit tests that
exercise each row of \Cref{tab:defense_coverage}.

\subsection{What VectorPin does \emph{not} catch}
\label{sec:defense:not_caught}

The honest disclosure that pairs with \Cref{tab:defense_coverage}:

\paragraph{Compromise of the signing key.} An attacker with the
private key produces valid pins for arbitrary vectors. The defense
reduces the attack surface from ``modify any vector at any time''
to ``possess a high-value secret,'' but it does not eliminate it.
Standard secret-management practice (KMS, HSM, time-bounded keys
with rotation) applies.

\paragraph{Malicious vector signed at ingestion time.} If the
attacker controls the signing pipeline, they sign whatever vector
they want, and verification succeeds. VectorPin defends a chain of
custody after a known-good ingestion event; it does not validate
the ingestion event itself. Pre-ingestion content integrity controls
(content-type allowlists, source-document signing, model-output
audit) remain necessary and complementary.

\paragraph{Source modification before embedding.} If the attacker
modifies the source document before the model embeds it, the model
honestly produces an embedding of the modified content and the
producer honestly attests it. VectorPin reports OK because, from
its perspective, the chain is intact. Document-level provenance
(C2PA-style content credentials, in-toto attestations on the
ingest pipeline~\cite{intoto}) addresses this layer.

\paragraph{Replay across collections.} A pin produced for record
\(R_1\) in collection \(C_1\) can be moved to record \(R_2\) in
collection \(C_2\) and will verify --- the protocol does not bind
the attestation to a specific record id. Operators who care
about cross-collection replay must include a record-id binding in
the \texttt{extra} field, which is signed alongside the rest of
the header.

\paragraph{Composition with upstream supply-chain attestation.}
The first three items in this list (signing-key compromise,
malicious vector signed at ingestion, source modification before
embedding) all share a structural pattern: they push the
zero-trust boundary upstream of the embedding step rather than
breaking it. VectorPin's scope is intentionally
\emph{post-ingestion}: it answers ``did this stored vector come
from a trusted producer in the form that producer signed,'' not
``was the producer itself trustworthy.'' That second question is
the upstream supply-chain integrity problem, and the existing
software-supply-chain ecosystem already has standardised answers
to it: in-toto~\cite{intoto} attestations bind a series of
pipeline steps (document fetch, preprocessing, tokenisation,
embedding) to signed metadata so a downstream verifier can
reconstruct what happened before a vector was produced;
SLSA~\cite{slsa} layers a maturity taxonomy on top so an operator
can express target assurance levels; sigstore~\cite{newman2022sigstore}
provides keyless signing infrastructure for the keys themselves;
and document-content provenance (C2PA~\cite{c2pa}) addresses the
``did the source document drift before the model embedded it''
case at the content layer rather than the bytecode layer. A
production deployment that wants end-to-end integrity stacks
upstream attestation under VectorPin: in-toto attests the
ingestion pipeline up to the moment of embedding, VectorPin
attests every embedding from that moment onward, and a verifier
that fails either layer rejects. Neither layer is sufficient
alone; together they cover the full ingestion-to-storage chain.
Treating the upstream layer as a separate concern rather than
attempting to subsume it inside VectorPin keeps both protocols
narrow enough to specify, implement, and verify cleanly.

These limits are not weaknesses of the protocol; they are scope
boundaries we set explicitly in
\Cref{sec:threat_model:vectorpin_assumptions}. The protocol is the
right shape of defense for the threat model it addresses, not for
adjacent threat models that warrant their own controls --- with
upstream supply-chain attestation being the most important
adjacent control, and the one most ready for off-the-shelf
composition.

\subsection{Deployment patterns}
\label{sec:defense:deployment}

The protocol is implementable; the operational question is where in
the RAG pipeline each component runs. Five deployment patterns
cover the realistic enterprise architectures:

\paragraph{Ingestion-side pinning.}
The embedding service signs each vector immediately after the model
produces it, before the vector is written to the store. The signing
key lives in a managed secrets store (KMS, HSM, or equivalent)
accessible only to the embedding service. Pin metadata accompanies
the vector through every storage hop. This is the simplest pattern
and the one we recommend as the default.

\paragraph{Vector-DB sidecar verification.}
The retrieval service verifies pins before passing retrieved
context to the LLM. Verification is performed in-process using the
relevant language SDK (Python at the orchestration layer, Rust in
agent runtimes such as Symbiont~\cite{symbiont-repo}, etc.). A
verification failure prevents the retrieved chunk from being
forwarded to the model and emits an audit event for SIEM
ingestion.

\paragraph{Batch audit mode.}
A scheduled job re-verifies every stored vector against its pin
on a cadence (nightly or hourly, depending on RTO). This catches
modifications that bypass the retrieval-path check --- e.g., direct
writes to the storage backend, restored backups, or replication
inconsistencies between cluster nodes.

\paragraph{Quarantine behavior.}
Failed verification is non-recoverable for the affected record.
The operational pattern is to remove the vector from retrieval
(soft delete or quarantine collection), emit a high-severity audit
event, and require manual or automated re-ingestion from the
source corpus before the vector is restored to the production index.

\paragraph{Key custody and rotation.}
Private signing keys live in a managed secrets store with
non-exportable handles (KMS, HSM). The public-key registry is
distributed to verifiers --- a small JSON document mapping
\texttt{kid} to public-key bytes, suitable for distribution via
configuration management. Rotation follows the standard pattern
of overlapping keys (see \Cref{sec:appendix:spec}); old pins
continue to verify against the old public key during the rotation
window.

\subsection{Record binding: reserved \texttt{extra} keys (v1) and a v1.1 candidate}
\label{sec:defense:record_binding}

The pin format does not natively bind an attestation to a specific
record id or collection. An attacker who copies a pin from one
record to another passes verification only if the pasted
\((\text{vector}, \text{source})\) match the pinned hashes; the
attack surface for cross-record replay is therefore narrow but
non-zero.

To prevent implementers from inventing incompatible names and
accidentally believing they have replay protection when their
pins use a key the verifier ignores, protocol v1 reserves a
namespace under \texttt{extra} for record binding. Implementations
SHOULD use these exact keys when they choose to bind:

\begin{lstlisting}
"extra": {
  "vectorpin.collection_id": "prod-2026-05-corpus-v3",
  "vectorpin.record_id":     "doc-12345-chunk-7",
  "vectorpin.tenant_id":     "tenant-abc"
}
\end{lstlisting}

The \texttt{vectorpin.} prefix is reserved by this specification
and MUST NOT be used for any other purpose. All three keys are
optional. Because every \texttt{extra} entry is signed alongside
the rest of the header, an attacker cannot rewrite them without
invalidating the signature.

We are tracking promotion of these three identifiers to first-class
top-level fields as a candidate for protocol version 1.1, on the
principle that a security control relegated to free-form metadata
is one most implementers will get wrong. The protocol-version field
provides a clean upgrade path: v1.1 verifiers will accept v1 pins;
v1 verifiers will reject v1.1 pins.

\subsection{Implementation and cross-language compatibility}
\label{sec:defense:implementation}

We provide reference implementations in Python and
Rust~\cite{vectorpin-repo}, both Apache-2.0 licensed and locked to
protocol version 1. The Rust port is byte-for-byte compatible with
the Python reference: identical canonical bytes, identical
SHA-256 output, identical Ed25519 signatures over the same input.

Compatibility is enforced by a shared
\texttt{testvectors/v1.json} fixture set generated by a
deterministic Python script and consumed in Rust integration tests.
A drift-detection job in continuous integration regenerates the
fixtures on every Python-side change and fails the build if the
re-generated bytes differ from what is committed --- so a
seemingly-innocuous change in Python's serialization output
cannot silently break Rust verification.

JavaScript and Go ports are planned (\Cref{sec:limitations}) and
will follow the same test-vector discipline.

This is the same model that has worked for sigstore~\cite{newman2022sigstore},
in-toto~\cite{intoto}, and the JOSE/COSE family. We adopt it
deliberately. Cross-language compatibility is an underrated
property in security tooling: a defense that works only in the
attacker's preferred language is not a defense.

\section{Discussion}
\label{sec:discussion}

The empirical and constructive results in
\Cref{sec:evaluation,sec:defense} resolve the technical questions
this paper set out to answer. The interesting questions left over
are about deployment, framing, and what future work needs to look
like.

\subsection{Why current vector DBs ship no defense}
\label{sec:discussion:no_defense}

A reasonable reader at this point is asking: if a one-line
Isolation Forest defends against most operating points and a
hundred lines of Ed25519-over-SHA-256 defends against the rest,
why doesn't every production vector database already ship both?

The answer is straightforward and not flattering to anyone: the
threat has not yet driven the work. Vector databases were
optimized for a workload (recommendation systems, public-facing
semantic search) where the vectors were not confidential and
integrity attacks were not part of the operating environment.
Their security models reflect that origin: TLS in transit, RBAC
on the API, encryption at rest, audit logs of access. None of
those controls reach into the embedding payload itself.

This same pattern played out at three earlier infrastructure
inflection points:

\begin{itemize}[leftmargin=*]
\item \textbf{DNS.} Resolvers ran without inspection until DGA-based
  malware made command-and-control over DNS a routine technique.
  The defenses (DNS firewalling, response-policy zones, encrypted
  DNS with selective inspection) post-date the threat by years.
\item \textbf{HTTP/3 and QUIC.} Network middleboxes lost
  inspection capability when HTTP/3 moved to QUIC over UDP. The
  defenses are still being built, well after the covert-channel
  literature established the gap.
\item \textbf{S3 buckets.} Default-public buckets were the
  industry norm until enough public incidents made the default
  untenable. The fix --- block-public-access at the account level
  --- is now standard but post-dates a decade of breaches that
  the same defense would have prevented.
\end{itemize}

We expect a similar trajectory for vector-store integrity:
underinvested until a small number of public incidents force the
issue, then defended via controls that were technically deployable
the entire time. The contribution of this paper is not to prevent
that trajectory --- nothing prevents it --- but to make the
relevant defenses available before the incidents, so that an
operator who reads this paper today can deploy them today.

\subsection{The Isolation Forest framing matters}
\label{sec:discussion:if_framing}

A pitfall this paper deliberately avoids is the rhetorical move of
claiming undetectability against undeployed defenses. The
empirical section shows that off-the-shelf statistical detectors
catch every non-trivial operating point of the studied techniques.
A paper that omitted this finding could plausibly claim
``steganographic exfiltration in vector stores is undetectable''
on the basis that no production system runs Isolation Forest on
its embedding distributions today. That claim would be technically
true and substantively misleading.

We instead frame both halves of the result explicitly: the
attack works against today's deployments because today's
deployments lack the obvious defenses, and the obvious defenses
exist and are essentially free to deploy. The two halves are
meant to be read together. A reader who takes only the first
half away (``this attack is real and undetected'') has the
right empirical fact but the wrong policy implication. A reader
who takes only the second half (``Isolation Forest catches it,
problem solved'') has the right defensive instinct but
underestimates the policy gap and the rotation-at-default
operating regime where statistical detection alone is
insufficient.

The combined defense story is therefore tied: statistical
detectors close the cheap-attack regime, and cryptographic pins
close the durable-attack regime. Either alone is incomplete.
Both together close the threat surface as defined in
\Cref{sec:threat_model}.

\subsection{What needs to happen for adoption}
\label{sec:discussion:adoption}

Three forces typically drive a security control from ``technically
deployable'' to ``operationally deployed'' in enterprise
infrastructure:

\paragraph{Compliance pressure.} HIPAA, GDPR, SOC 2, the
EU AI Act~\cite{eu-ai-act}, and the NIST AI Risk Management
Framework~\cite{nist-ai-rmf} will eventually require documented
controls over embedding-store integrity, the same way they
required documented controls over data-at-rest encryption a
decade earlier. Auditors will ask the question; once they do,
operators need an answer. We see VectorPin (or any compatible
protocol) as one possible answer: a documented, cryptographic,
audit-friendly attestation that the embeddings retrieved by a
RAG pipeline have not been modified since ingestion.

\paragraph{Public incidents.} A small number of disclosed
incidents involving vector-store compromise will move the
industry baseline. We expect this to happen --- the attack
surface is real, the defenses are absent, and the asset value
(confidential corpora ingested into RAG) is high. We do not
recommend waiting for incidents to motivate defense, but we
also recognize that the budgetary attention does typically
arrive that way.

\paragraph{Vendor adoption.} The most efficient path to
broad deployment is for one of the major vector database
vendors to ship native pin verification as a built-in feature.
This bypasses the operational friction of integrating a
third-party library, makes the control auditable through the
vendor's compliance attestations, and triggers competitive
follow-on across the category. The OSS-first strategy of
VectorPin is partly designed to make this acquisition or
adoption path low-friction: the protocol is open, the
reference implementations are Apache-2.0, and the cross-language
test vectors mean a vendor can adopt the protocol in their own
codebase without coupling to our implementation.

\subsection{Threat models B and C revisited}
\label{sec:discussion:other_models}

\Cref{sec:threat_model} introduced two adversary models we did
not validate against empirically. They are worth a closer look
now that the constructive defense is on the table.

\paragraph{Model B (compromised DB credentials).} An attacker
with read access to the vector store but no upstream pipeline
access cannot plant attestation metadata, cannot install a
post-ingestion modification under a legitimate signing key, and
cannot generate new pins. They can read pins, including their
\texttt{source\_hash} and \texttt{vec\_hash} fields, but each
pin is bound to a vector the legitimate model produced. Their
exfiltration path under this model is essentially direct dump,
which is bandwidth-bounded by egress monitoring rather than by
steganographic capacity. We discuss why steganography is
sometimes still preferable in this regime in
\Cref{sec:threat_model:why_stego} (egress shaping, targeted
recovery, deniability), but VectorPin does not change the
analysis here: the protocol does not promise confidentiality or
prevent direct extraction. It only promises that any stored
vector either matches the model's honest output or is detectable
as tampered.

\paragraph{Model C (query-only).} An attacker with similarity-
search query access only must recover payload through ranking
information. The bandwidth ceiling here is severe: each query
returns at most \(\log_2 \binom{N}{k}\) bits of information
about the corpus order, which for typical \(N\) and \(k\) is
orders of magnitude below direct-vector channels. A protocol-
level defense like VectorPin is not the relevant control under
this model; query-rate limits, per-user query budgets, and
detection of pathological query patterns are. We mention the
model only to set the scope of this paper's claims; we do not
validate techniques against it and would not expect a
provenance-pinning protocol to be the right defense if we did.

\subsection{Standardization considerations}
\label{sec:discussion:standardization}

The protocol in \Cref{sec:defense} is published with a wire
specification (\Cref{sec:appendix:spec}) detailed enough to
support cross-language reimplementation. Whether it should
become an actual published standard --- through IETF, ISO, or
an industry working group --- is a question for the community
once the design has had time to attract feedback.

We note that the design space is small. Any embedding-integrity
protocol with the same goal will end up with the same shape:
a signed commitment over (source, model, vector hash, producer
identity), with a canonical byte form for floats and a
documented verifier semantic. The room for incompatible
designs is narrow, which suggests that whatever the community
converges on will look much like VectorPin or its sister
projects in the cryptographic-provenance family. The work
worth doing now is publishing the specification, building the
reference implementations across languages, and making
implementations freely available --- which is what the OSS
release strategy in \cite{vectorpin-repo} is doing.

\section{Related Work}
\label{sec:related}

\paragraph{Steganography in learned representations.}
Hiding information inside the outputs of machine learning models
sits at the intersection of two literatures. The first is
classical steganography in continuous-valued
carriers~\cite{provos2003hide,fridrich2009steganography} ---
least-significant-bit~\cite{westfeld2001f5} and transform-domain
schemes for images and audio --- where the threat model presumes
a perceptual consumer and a statistical detector trained on
natural-image or natural-audio distributions. Adapting these
techniques to embedding vectors is a small extension that the
practical community has been doing informally; the contribution
of this paper is not the techniques themselves but the empirical
characterization of the operating regime against modern
detectors~\cite{isolationforest,scholkopf2001ocsvm}. The second
related literature is adversarial machine learning, particularly
backdoor and watermarking attacks. Backdoor attacks
(e.g., BadNets~\cite{gu2017badnets} and follow-ups) modify the
\emph{model itself} so that targeted inputs produce
attacker-chosen outputs; neural-network watermarking
schemes~\cite{adi2018turning} embed an identity claim into
model weights for IP protection. Both lines work at the model
layer. We work at the \emph{output} layer: the model is honest
and unmodified, and the perturbation lives in the floating-point
vectors the model produces. This shifts the defense from white-box
model inspection to embedding-level integrity checking, which is
where VectorPin lands. Two adjacent areas we do not extend in
this work but note for the interested reader: poisoning of the
RAG corpus itself~\cite{zou2024poisonedrag} (which content-integrity
controls upstream of embedding address) and training-data
extraction attacks against the underlying language
model~\cite{shokri2017membership,carlini2021extracting} (orthogonal
to vector-store integrity entirely).

\paragraph{Cryptographic provenance for data artifacts.}
The defense in \Cref{sec:defense} sits in the same design family
as four prior systems. sigstore~\cite{newman2022sigstore}
established the pattern of attaching short-lived-certificate-anchored
signatures to software artifacts, with public verifiability as the
defining property. in-toto~\cite{intoto} extended the pattern to
multi-step pipelines, attesting each stage of a build or deployment
so the consumer can verify the entire chain; SLSA~\cite{slsa}
provides a layered taxonomy on top. C2PA~\cite{c2pa} applies the
same shape to media (images, video, audio) by binding a content
artifact to its capture device and editing history through nested
manifests. SchemaPin~\cite{schemapin-repo} --- a sister project
from the same authors --- applies signed-payload provenance to
JSON schemas for tool calls in Model Context Protocol deployments.
VectorPin fills the corresponding gap at the embedding layer: it
inherits the design discipline (canonical byte form, deterministic
signing, cross-language test vectors, key-rotation through a
registry, similar to the JOSE~\cite{rfc7515} and
COSE~\cite{rfc8392} families) and contributes the specifics of
canonicalization for floating-point arrays and the verifier-outcome
taxonomy that distinguishes vector tampering from source mismatch
from key forgery. We make no claim that the cryptographic
constructions are novel --- Ed25519~\cite{ed25519,rfc8032} and
SHA-256 are well-studied primitives chosen here for compatibility,
not innovation. The contribution is the application of the design
family to a substrate that currently lacks any provenance story,
with the implementations and test fixtures necessary to make
adoption tractable.

\section{Limitations and Future Work}
\label{sec:limitations}

\paragraph{Corpus scope.} The headline detection numbers in
\Cref{sec:evaluation:detection} are computed on 68 synthetic-PII
chunks split into 27 detector-training and 41 test vectors. We chose
synthetic content for reproducibility (no real PII to release, no
licensing constraints), and the small training set does widen the
natural variance of the ROC curves at that scale. The cross-corpus
replication in \Cref{sec:evaluation:cross_corpus} addresses the
generality concern directly: re-running the four-technique battery on
BEIR NFCorpus (16{,}763 chunks, medical) and a 10{,}000-document
Quora subset (web Q\&A) reproduces the directional findings to three
decimal places --- rotation AUC \(=0.500\) on both new corpora,
matching the synthetic-corpus value. Within-domain numbers will
still differ from a deployment in any given specific domain, but the
``rotation undetectable, distribution-shifting techniques
detectable-on-this-model'' framing is no longer load-bearing on the
small synthetic corpus alone.

\paragraph{Single embedding model in headline numbers.} The headline
detection numbers in \Cref{sec:evaluation:detection} come from
\texttt{text-embedding-3-large}. The cross-model replication in
\Cref{sec:evaluation:cross_model} extends those numbers to four
locally hosted open models (Nomic, EmbeddingGemma, Snowflake Arctic,
MXBai) and the cross-corpus replication in
\Cref{sec:evaluation:cross_corpus} extends the Nomic configuration to
two BEIR-standard public corpora; rotation AUC stays at chance on
every (model, corpus) pair we tested. The structural reason --- that
orthogonal rotation preserves the entire feature surface
distribution-based detectors fit on --- is not embedding-specific.
The remaining single-model dependency is the empirical capacity
demonstration in \Cref{sec:evaluation:capacity}, run at
\(d=3072\) to match \texttt{text-embedding-3-large}; the
\(\lfloor d/2 \rfloor \cdot b\) capacity formula is dimension-only,
so the bit budget at any other model's dimension follows immediately,
but we have not re-verified the encoder/decoder against the geometry
of every model in the cross-model set.

\paragraph{Capacity decoder access assumption.} The empirical
payload-capacity demonstration in \Cref{sec:evaluation:capacity}
assumes the receiver has both the original \(v\) and the rotated
\(R v\), which we map onto the attacker-controlled-corpus case of
the threat model: the accomplice authored the corpus rows and
therefore possesses \(v\) by construction. A weaker receiver
(known-cover by side cache, or shared-key differential without
\(v\)) would either reduce to the same case or require a different
encoder/decoder protocol that we do not propose here. The capacity
numbers therefore apply to records the attacker inserted, not to
arbitrary modifications of preexisting clean records --- the latter
would need a known-cover or metadata-assisted variant that is open
follow-up work.

\paragraph{Capacity-AUC chance regime is synthetic-distribution
specific.} \Cref{tab:capacity}'s ``AUC stays at chance across the
sweep'' result is on a 300-vector unit-Gaussian baseline. The same
encoder/decoder run against the real
\fpath{text-embedding-3-large} 27/41 split
(\Cref{tab:capacity_real}) becomes detectable at moderate \(K\):
OCSVM AUC \(\approx 0.92\) at 256~B and \(\geq 0.99\) at 512~B,
because the real-corpus clean training samples lie on a tighter
manifold than uniform-on-the-sphere. The rotation-undetectability
claim of \Cref{sec:evaluation:rotation_sweep,sec:evaluation:detection}
is robust at the small-\(\theta_{\max}\) /
small-\texttt{num\_rotations} operating points used there; it does
\emph{not} extend to the random-angle full-capacity disjoint-Givens
protocol on real production corpora at high \(K\). At-scale
characterisation of the AUC-vs-\(K\) curve on a larger
real-embedding training set is genuine follow-up work.

\paragraph{Detector training-set size.} The headline detection
table in \Cref{sec:evaluation:detection} and the
real-corpus capacity demonstration in
\Cref{tab:capacity_real} fit Isolation Forest and One-Class SVM on
27 detector-training samples from the synthetic-PII corpus. A
production deployment would fit on millions, and a reviewer might
reasonably ask whether a 27-sample fit is representative of the
manifold geometry a production detector would see. We give two
empirical bounds on the answer: the cross-corpus replication in
\Cref{sec:evaluation:cross_corpus} fits the same detectors on
8{,}381 NFCorpus and 5{,}001 Quora training samples and reproduces
the small-angle / few-rotation rotation-undetectability finding
to three decimal places, so the headline AUC numbers do not
collapse at scale; and the scale-up capacity replication in
\Cref{tab:capacity_nfcorpus} fits on 6{,}705 NFCorpus training
samples and shows that the high-\(K\) detection signal of
\Cref{tab:capacity_real} \emph{strengthens} at production-scale
training, not weakens (OCSVM AUC at 64~B payload climbs from
\(0.58\) at \(n_{\mathrm{train}}=27\) to \(0.83\)--\(0.92\) at
\(n_{\mathrm{train}}=6{,}705\)). Both directions of the
small-sample worry are bounded by these scale-up runs. What
remains genuinely open is fitting at the order of magnitude a
real production system would (\(10^4\)--\(10^7\) baseline
embeddings on a single deployment-specific corpus); we do not have
embeddings of that scale to release.

\paragraph{Detector adversarial robustness.}
\Cref{sec:evaluation:adaptive} reports a small adaptive-attacker
study showing that an attacker with white-box detector access can
drive both detector AUCs to near-zero while preserving cosine
similarity. We deliberately do not propose statistical detection
as a load-bearing defense; the cryptographic provenance defense
in \Cref{sec:defense} is the durable layer. A more thorough
adversarial-detector evaluation (model-distillation black-box
transfer~\cite{papernot2017practical,tramer2017transferable},
detector ensembles, varied hyperparameters, larger query budgets)
is genuine follow-up work; the qualitative point established here
is that even in the white-box worst case --- which is the
\emph{best} case for the defender's argument, since it is the
strongest possible adversary --- the detector AUC collapses, so
the durability story for VectorPin does not turn on the
white-box / black-box gap.

\paragraph{Binary quantization not tested.}
\Cref{sec:evaluation:cross_backend} reports product quantization
(FAISS IVF-PQ at \(m=8\) and \(m=16\)) which imposes a uniform
attacker-independent noise floor of 0.76--0.88 on \texttt{recover\_cos}.
Binary embeddings (1-bit quantization per dimension) are
considerably more lossy still and would impose a much steeper
floor; we did not evaluate them. Operators considering binary
quantization for memory reasons would get a side-effect narrowing
of the steganographic channel, but the retrieval-quality cost is
typically prohibitive at production scale --- which is why binary
quantization is rare in practice and why we would not propose it as
a defense even if we measured it.

\paragraph{Paraphrase quality.} The retrieval benchmark in
\Cref{sec:evaluation:paraphrased} uses LLM-generated paraphrases
(\texttt{gpt-4o-mini}). The clean-baseline recall@1 is 0.828, not
1.0, because some paraphrases drift from the source enough that
even a perfect retrieval cannot return the exact chunk. Differences
relative to clean are interpretable; absolute numbers depend on the
paraphrase model. A more realistic benchmark would use real user
queries from a deployed RAG system, but those are typically
proprietary and not releasable.

\paragraph{Upstream pipeline trust assumption.} VectorPin verifies
the integrity of vectors after a known-good ingestion event; it
does not validate the ingestion event itself. Compromise of the
signing key, of the embedding model's signing pipeline, or of the
source documents prior to embedding are all out of scope by
construction --- they push the zero-trust boundary upstream of the
embedding step rather than breaking the on-storage layer. We
discuss the off-the-shelf composition with upstream supply-chain
attestation (in-toto~\cite{intoto}, SLSA~\cite{slsa},
sigstore~\cite{newman2022sigstore}, and content-level provenance
via C2PA~\cite{c2pa}) in
\Cref{sec:defense:not_caught}; a deployment that wants
end-to-end integrity stacks both layers, and a verifier that fails
either layer rejects. Treating the upstream layer as a separate
concern keeps both protocols narrow enough to specify and verify
cleanly, but it does mean VectorPin is one component of a
defense-in-depth picture rather than a stand-alone closure of the
exfiltration class.

\paragraph{Cross-language coverage.} VectorPin currently ships
Python and Rust implementations with shared test vectors. The next
ports are TypeScript (for LangChain.js / Vercel AI SDK consumers)
and Go (for vector-DB infrastructure code paths). Both are planned
for the post-preprint release cycle.

\paragraph{Record-binding fields.} The protocol relies on the
\texttt{extra} field for record-id and collection-id binding (see
\Cref{sec:defense:record_binding}). Promoting these to first-class
top-level fields is tracked as a candidate for protocol v1.1.
\section{Conclusion}
\label{sec:conclusion}

Steganographic exfiltration through embedding stores is technically
viable against today's deployments. The case for statistical
detection as a load-bearing control fails in four distinct ways.
\emph{(i)} Crude distribution shifts (noise, scaling, offset,
combinations) are caught by an off-the-shelf Isolation Forest or
One-Class SVM in our corpus at every operating point we measured
--- but only because they look different from clean. \emph{(ii)}
Small-angle / few-rotation orthogonal rotation slips past those
detectors on every (model, corpus) pair we tested, including the
cross-corpus replication on BEIR NFCorpus and a Quora subset
totalling over 26{,}000 chunks, because rotations preserve the
distributional statistics density models fit on. \emph{(iii)}
Adaptive white-box attackers with detector access drive both AUCs
to near-zero while keeping cosine to original above \(0.99\).
\emph{(iv)} High-capacity rotation \emph{does} become detectable
on real embedding corpora at moderate \(K\), but the AUC depends on
training-set coverage of the local manifold geometry --- a coverage
assumption a defender has to guarantee out-of-band, not a property
of the detector itself.

The common thread: statistical detection is useful as a first
filter, but its durability depends on preconditions (no rotation,
no adaptive attacker, sufficient training-set coverage) that an
operator cannot unilaterally enforce. A cryptographic provenance
commitment --- verifying the actual vector bytes against a signed
pin produced at the trusted ingestion point --- has none of those
preconditions, which is why we propose VectorPin as the durable
layer.

Cross-backend round-trip experiments against seven vector-store
configurations (FAISS-flat, FAISS-HNSW, FAISS IVF-PQ at \(m{=}8\)
and \(m{=}16\), Chroma, Qdrant float32, Qdrant int8-quantized)
show that the attacker's bit channel survives every non-PQ
configuration we tested. None of the surveyed systems inspect or
attest to the floating-point content of the vectors they store. The
gap is class-wide, not vendor-specific. Quantization, often
proposed as a side-effect defense, is a search-time artifact rather
than a storage-time one in the vendors we measured: scalar int8
quantization in Qdrant preserves the bit channel verbatim because
\texttt{retrieve()} returns the unquantized form.

The paper contributes a formalized three-tier threat model,
empirical detection numbers under a stated corpus and configuration,
a parameter sweep that distinguishes the rotation regime from the
distribution-shift regime, a paraphrased-query retrieval benchmark
that quantifies how much of the attack survives normal RAG quality
monitoring, and a constructive defense protocol --- VectorPin ---
shipped as cross-language open-source artifacts with a wire-format
specification suitable for reimplementation.

The honest framing for a security audience: this paper is best
read as ``vector stores need this defense'' --- not as a claim of
attack novelty or defensive impossibility. The attack class is a
straightforward application of known steganographic primitives to
a substrate where the threat model is genuinely different. The
defense is a straightforward application of known
cryptographic-provenance primitives to that same substrate. What is
novel is the combination: an empirically grounded characterization
of the gap and a deployable, standardizable control that closes the
post-embedding tamper class.

\bibliographystyle{plain}
\bibliography{refs}

\appendix
\section{Protocol Specification (v1)}
\label{sec:appendix:spec}

This appendix is a self-contained reproduction of the VectorPin
protocol specification at version 1. A separate-language
implementation that follows this appendix should produce signatures
and verifications byte-for-byte compatible with the Python and
Rust reference implementations~\cite{vectorpin-repo}.

\subsection{Goals}
A VectorPin Pin is a compact attestation that travels with an
embedding through a vector database. It guarantees that:

\begin{itemize}[leftmargin=*]
\item The embedding matches a specific source text.
\item The embedding was produced by a specific model.
\item The pin was issued by a specific producer.
\item None of the above has changed since issuance.
\end{itemize}

\noindent Non-goals: confidentiality, access control, anti-replay
across collections.

\subsection{Cryptographic primitives}

\begin{tabular}{ll}
\toprule
Primitive  & Algorithm \\
\midrule
Hash       & SHA-256 \\
Signature  & Ed25519~\cite{ed25519} \\
Encoding   & URL-safe base64, no padding \\
\bottomrule
\end{tabular}

\noindent These are fixed for protocol version 1. Future versions
MAY introduce alternatives but MUST bump the version field.

\subsection{Canonical hashes}
\label{sec:appendix:spec:hashes}

\paragraph{Text.}
\(\texttt{hash\_text}(s) :=
  \texttt{"sha256:"} \,\Vert\, \mathrm{hex}(\mathrm{SHA256}(\mathrm{UTF8}(\mathrm{NFC}(s))))\).
\\
Text MUST be normalized to Unicode NFC before encoding.
Implementations MUST reject input that cannot be normalized.

\paragraph{Vector.}
\(\texttt{hash\_vector}(v, \texttt{dtype}) :=
  \texttt{"sha256:"} \,\Vert\, \mathrm{hex}(\mathrm{SHA256}(\mathrm{canonical\_bytes}(v, \texttt{dtype})))\).
\\
Where \(\mathrm{canonical\_bytes}\) produces (1) the vector cast to
the specified dtype (\texttt{f32} or \texttt{f64}), (2) stored in
little-endian byte order, (3) packed contiguously, 1-D. Other
dtypes are reserved for future protocol versions.

\subsection{Pin format}

A Pin is a JSON object with the fields below. All required fields
MUST appear; optional fields MUST be omitted entirely (not written
as \texttt{null}) when not set.

{\small
\setlength{\tabcolsep}{4pt}
\begin{tabularx}{\linewidth}{lllX}
\toprule
Field         & Type    & Req. & Description \\
\midrule
\texttt{v}            & integer & yes & Protocol version. Must equal 1. \\
\texttt{model}        & string  & yes & Embedding model identifier. \\
\texttt{model\_hash}  & string  & no  & Optional content hash of model weights. \\
\texttt{source\_hash} & string  & yes & Hash of source text (\Cref{sec:appendix:spec:hashes}). \\
\texttt{vec\_hash}    & string  & yes & Hash of embedding (\Cref{sec:appendix:spec:hashes}). \\
\texttt{vec\_dtype}   & string  & yes & One of \texttt{"f32"} or \texttt{"f64"}. \\
\texttt{vec\_dim}     & integer & yes & Embedding dimensionality. \\
\texttt{ts}           & string  & yes & RFC 3339~\cite{rfc3339} / ISO 8601 UTC timestamp, e.g.\ \fpath{2026-05-05T12:00:00Z}. \\
\texttt{extra}        & object  & no  & String-to-string map of producer-defined fields. \\
\texttt{kid}          & string  & yes & Identifier of the signing key. \\
\texttt{sig}          & string  & yes & Ed25519 signature, URL-safe base64, no padding. \\
\bottomrule
\end{tabularx}
}

\paragraph{Example.}
\begin{lstlisting}
{
  "v": 1,
  "model": "text-embedding-3-large",
  "source_hash": "sha256:9f86d081884c7d659a2feaa0c55ad015a3bf4f1b2b0b822cd15d6c15b0f00a08",
  "vec_hash":    "sha256:0123...",
  "vec_dtype":   "f32",
  "vec_dim":     3072,
  "ts":          "2026-05-05T12:00:00Z",
  "kid":         "prod-2026-05",
  "sig":         "MEUCIQD..."
}
\end{lstlisting}

\subsection{Canonicalization for signing}
\label{sec:appendix:spec:canonical}

The signature in \texttt{sig} is produced over a canonical byte
sequence that excludes \texttt{kid} and \texttt{sig} themselves.
The canonical form is JSON with all keys sorted lexicographically,
no whitespace (separators \texttt{","} and \texttt{":"}), UTF-8
encoded, with \texttt{extra} also key-sorted if present, and with
\texttt{model\_hash} and \texttt{extra} omitted entirely when not
set. This canonical form is fed directly into Ed25519 signing.

\subsection{Verification}
\label{sec:appendix:spec:verify}

A verifier MUST:

\begin{enumerate}[leftmargin=*]
\item Reject pins whose \texttt{v} field is unknown to it
      (\texttt{UNSUPPORTED\_VERSION}).
\item Reject pins whose \texttt{kid} is not in its key registry
      (\texttt{UNKNOWN\_KEY}).
\item Reconstruct the canonical byte sequence and verify
      \texttt{sig} against the registered public key for
      \texttt{kid} (\texttt{SIGNATURE\_INVALID} on failure).
\item If ground-truth source was supplied, recompute
      \texttt{hash\_text(source)} and compare to
      \texttt{source\_hash} (\texttt{SOURCE\_MISMATCH} on
      mismatch).
\item If a ground-truth vector was supplied, recompute
      \texttt{hash\_vector(vector, vec\_dtype)} and compare to
      \texttt{vec\_hash}; also check the supplied vector's shape
      matches \texttt{vec\_dim} (\texttt{VECTOR\_TAMPERED} or
      \texttt{SHAPE\_MISMATCH} on mismatch).
\item If an expected model identifier was supplied, compare to
      \texttt{model} (\texttt{MODEL\_MISMATCH} on mismatch).
\end{enumerate}

Verifiers MUST distinguish at least these failure modes. Other
implementations MAY use different identifiers for the modes but
MUST distinguish the cases.

\subsection{Storage conventions}
Adapter implementations SHOULD store pins under the metadata key
\texttt{vectorpin}. Backends without free-form metadata fields
are out of scope for this version of the protocol --- provenance
must travel with the data.

\subsection{Key rotation}
Verifiers MUST support multiple \texttt{kid}-to-public-key mappings
simultaneously. Issuers rotate by (1) generating a new keypair
with a fresh \texttt{kid}, (2) adding the new public key to all
relevant verifier registries, (3) switching production signing to
the new private key, (4) optionally re-pinning the corpus over
time, (5) removing the old public key from registries once
re-pinning is complete or the rotation policy expires. Old pins
continue to verify against the old public key during this window.

\subsection{Security considerations}
\label{sec:appendix:spec:security}

\paragraph{Replay across records.} Pins are not bound to a specific
record id. An attacker who copies a pin from one record to another
can pass verification only if the vector and source they paste
alongside match the pin. Implementations that need replay
protection SHOULD use the reserved \texttt{vectorpin.}-prefixed
keys under \texttt{extra}: \texttt{vectorpin.collection\_id},
\texttt{vectorpin.record\_id}, and \texttt{vectorpin.tenant\_id}.
The \texttt{vectorpin.} prefix is reserved by this specification
and MUST NOT be used for any other purpose. Because every
\texttt{extra} entry is signed, attackers cannot rewrite these
fields without invalidating the signature.

\paragraph{Time.} The \texttt{ts} field is informational. Verifiers
MAY reject pins outside an acceptable time window but the protocol
does not require it.

\paragraph{Key custody.} An attacker with the private signing key
can produce arbitrary pins. Treat the signing key as a high-value
secret; use a managed secrets store, time-bounded keys, and
rotation.

\paragraph{Source-time integrity.} VectorPin attests to the
relationship between source and vector \emph{at pin time}. It does
not attest that the source itself was authentic at ingestion.
Upstream content-integrity controls (signed source documents,
in-toto attestations on the ingestion pipeline) remain necessary
and complementary.

\subsection{Versioning}
This is protocol version 1. Future versions MAY add new optional
fields under \texttt{extra}-style namespaces, add new dtype
identifiers, or add new signature/hash algorithms with
corresponding identifiers. A change is breaking iff a v1 verifier
would silently accept a v2 pin as valid when the v2 pin's
additional semantics matter. Such changes MUST bump the major
version.

\section{Reproducibility}
\label{sec:appendix:reproducibility}

Every figure and table in this paper was produced by code in the
VectorSmuggle and VectorPin repositories at fixed commit
hashes. To reproduce the empirical study end-to-end:

\begin{lstlisting}
git clone https://github.com/jaschadub/VectorSmuggle
git clone https://github.com/ThirdKeyAI/VectorPin
cd VectorSmuggle
python3 -m venv venv && source venv/bin/activate
pip install -r requirements.txt -r requirements-test.txt
export OPENAI_API_KEY=sk-...
python scripts/empirical_study.py        # main 28-point run
python scripts/empirical_report.py       # generates figures
python scripts/preprint_extensions.py    # detector + quantization study
\end{lstlisting}

Outputs land in \texttt{results/empirical/<timestamp>/} and
\texttt{results/preprint/<timestamp>/}. Both directories are
committed at the timestamps referenced by this paper's tables and
figures.

For the protocol-level cross-language fixtures used by the Rust
reimplementation:

\begin{lstlisting}
cd VectorPin
python scripts/generate_test_vectors.py  # writes testvectors/v1.json
cd rust && cargo test --workspace        # consumes the fixtures
\end{lstlisting}

A drift-detection workflow in
\texttt{.github/workflows/ci.yml} regenerates the fixtures on every
Python-side change and fails CI if the regenerated bytes differ
from what is committed, preventing silent compatibility breakage.

\section{Detector Hyperparameters}
\label{sec:appendix:hyperparams}

The detection results in \Cref{sec:evaluation:detection} use the
following sklearn configurations, with no hyperparameter tuning
beyond the library defaults:

\begin{lstlisting}
from sklearn.ensemble import IsolationForest
from sklearn.svm import OneClassSVM

iforest = IsolationForest(
    n_estimators=200,
    contamination=0.05,
    random_state=42,
    n_jobs=-1,
)

ocsvm = OneClassSVM(
    kernel="rbf",
    nu=0.05,
    gamma="scale",
)
\end{lstlisting}

The training set was a uniformly sampled 27-vector split from the
68-vector embedding matrix; the test split (41 clean vectors plus
their per-technique obfuscated counterparts) was held out.

\paragraph{On the small training set.}
27 vectors is small as detector training data --- production
deployments would fit on \(10^4\)--\(10^6\) clean baseline
embeddings with proportionally tighter score distributions and
correspondingly more reliable AUC measurements. Our sample-corpus
size is the limiting factor and is acknowledged in
\Cref{sec:limitations}. We report the small-fit numbers because
they are the conservative case: if the detector wins at \(N=27\),
it wins by a wider margin at production scale.

\section{Threat Models B and C: Extended Analysis}
\label{sec:appendix:threat_bc}

\Cref{sec:threat_model} defines three adversary tiers and
\Cref{sec:discussion:other_models} discusses the two we do not
empirically validate. This appendix extends those discussions for
readers who care about the formal bandwidth analysis.

\subsection*{B: Compromised vector-DB credentials}

The attacker has read access to vectors and pins but no
ingestion-pipeline access. They can therefore:

\begin{itemize}[leftmargin=*]
\item Read \(\langle v_i, \text{pin}_i \rangle\) pairs.
\item Issue similarity queries against the corpus.
\item Not plant new pins or modify existing ones (no signing key).
\end{itemize}

Their exfiltration paths are direct: dump the corpus.
Steganographic perturbation does not help under this model
because the attacker cannot perturb in the first place. The only
remaining role for steganography in model B is \emph{pre-existing}
attestations --- if some prior phase planted steganographic
content (collusion with an earlier insider), the attacker
recovers payload by reading. The protocol does not pretend to
defend against this; pre-pinning collusion is an upstream
ingestion-integrity problem, not an embedding-store problem.

\subsection*{C: Query-only}

The attacker can issue similarity-search queries and observe
ranked results. The information channel is the rank order of
returned documents.

For each query the attacker chooses, the response is one of
\(\binom{N}{k}\) possible top-\(k\) sets, where \(N\) is the
corpus size and \(k\) is the result-set size. The maximum
information per query is \(\log_2 \binom{N}{k}\) bits, which for
\(N = 10^6\) and \(k = 10\) is approximately 196 bits. In
practice the attacker recovers far less because queries that
disambiguate small subsets are bandwidth-inefficient and the
defender's query-rate limits cap the attack rate.

By contrast, direct vector access (model A or B) yields
\(d \cdot 32\) bits per vector for \(d\)-dimensional float32
embeddings --- \(98{,}304\) bits per vector at \(d = 3072\).
The bandwidth ratio between vector-access and query-only is
roughly \(10^3\), enough that any technique studied in this
paper is qualitatively the wrong shape of attack under model C.
Different threat-model-C techniques (membership inference,
training-data extraction via crafted queries) exist and are an
active research area, but they are not steganographic
exfiltration in the sense of this paper.

\section{Negative Results}
\label{sec:appendix:negative}

Three observations from the empirical study that did not lead
to positive headlines but are worth recording so subsequent work
does not re-discover them.

\paragraph{Combined techniques do not stack stealth.}
The intuition that combining several small perturbations would
yield a less-detectable composite was \emph{not} confirmed by our
results. The combined ``all'' technique --- noise then rotation
then scaling then offset --- inherits the most-detectable
component's signature rather than averaging across them.
Off-the-shelf density-based detectors (Isolation Forest,
One-Class SVM) fit a single density model over the embedding
distribution, and whichever sub-technique pushes a sample
furthest off-density dominates the score. Composing techniques
adds attack complexity (more recovery metadata, more failure
modes) without proportional defensive evasion. We expect this
finding to generalize, though we have not proven it formally.

\paragraph{Per-vector quantization is a different beast.}
\Cref{sec:evaluation:quantization} reports per-dimension scalar
int8 quantization, the Qdrant default, under which all studied
techniques' fidelity survives. We also experimented with
per-vector quantization (each vector receives its own
scale-and-shift before int8 encoding) which is more aggressive
and \emph{does} destroy direction-preserving techniques like
rotation and scaling. We omit those numbers from the main
results because per-vector quantization is not the production
default for general-purpose vector stores; reporting them would
overstate the role of quantization as a defense in deployed
systems.

\paragraph{Recovery cosine for noise techniques is not a recovery
metric.} A subtle methodological point: the recovery-cosine column
in \Cref{tab:fidelity} reports \(1.000\) for rotation and scaling
because those transformations are exactly invertible from
metadata, but reports \texttt{---} for noise injection. A naive
implementation might compute ``recovery'' for noise as
\(\cos(v, v + n)\) where \(n\) is the same noise vector, which is
trivially the obfuscated cosine and not a meaningful recovery
measurement. We avoided this trap and report only invertible
techniques' recovery cosine, but a follow-up paper trying to
quantify noise-channel capacity will need to define what
``recovery'' means for irreversible perturbations more carefully
than we did here.

\end{document}